\documentclass[a4paper,fleqn,usenatbib]{mnras}
\usepackage{times}
\usepackage[dvipdfmx]{graphicx,pdflscape}
\usepackage{xspace}
\usepackage{amssymb}
\usepackage{amsmath}
\usepackage{multirow}

\newcommand{\apm}{APM~08279+5255\xspace}
\newcommand{\pds}{PDS~456\xspace}
\newcommand{\hhfull}{1H~0707-495\xspace}
\newcommand{\hh}{1H0707\xspace}
\newcommand{\xmm}{{\it XMM-Newton}\xspace}
\newcommand{\nustar}{{\it NuSTAR}\xspace}
\newcommand{\suzaku}{{\it Suzaku}\xspace}

\newcommand{\monaco}{{\sc monaco}\xspace}

\def\mnras{MNRAS}
\def\apj{ApJ}
\def\apjl{ApJL}
\def\apjs{ApJS}
\def\aap{A{\&}A}

\def\pasj{PASJ}

\title[A disk wind interpretation for \hhfull]
{A disk wind interpretation of the strong Fe K$\alpha$ features in \hhfull}
\author[K. Hagino et al.]  {
Kouichi Hagino$^{1}$, Hirokazu
Odaka$^{2}$, Chris Done$^3$, Ryota Tomaru$^{1,4}$, Shin Watanabe$^{1,4}$,
\newauthor Tadayuki Takahashi$^{1,4}$\\
$^1$ Institute of Space and Astronautical Science (ISAS), Japan Aerospace Exploration Agency (JAXA), 3-1-1 Yoshinodai, Chuo, Sagamihara,\\ Kanagawa 252-5210, Japan\\
$^2$ KIPAC, Stanford University, 452 Lomita Mall, Stanford, CA 94305, USA\\
$^3$ Centre for Extragalactic Astronomy, Department of Physics, University of Durham, South Road, Durham DH1 3LE, UK\\
$^4$ Department of Physics, University of Tokyo, 7-3-1 Hongo, Bunkyo, Tokyo 113-0033, Japan\\
}

\date{Submitted to MNRAS}
\pubyear{}

\def\msun{M$_\odot$\xspace}
\def\apj{ApJ}

\begin{document}
\label{firstpage}
\pagerange{\pageref{firstpage}--\pageref{lastpage}}
\maketitle

\begin{abstract}
\hhfull is the most convincing example of a supermassive black hole
with an X-ray spectrum being dominated by extremely smeared,
relativistic reflection, with the additional requirement of strongly supersoler iron abundance. 
However, here we show that the iron features
in its 2--10~keV spectrum are rather similar to the archetypal wind
dominated source, \pds. We fit all the 2--10~keV spectra from
\hhfull using the same wind model as used for \pds, but viewed
at higher inclination so that the iron absorption line is broader but
not so blueshifted. This gives a good overall fit to the data from
\hhfull, and an extrapolation of this model to higher energies also
gives a good match to the \nustar data. Small remaining residuals
indicate that the iron line emission is stronger than in \pds. This is consistent
with the wider angle wind expected from a
continuum driven wind from the super-Eddington mass accretion
rate in \hhfull, and/or the presence of residual reflection from the underlying disk
though the presence of the absorption line in the model 
removes the requirement for highly relativistic smearing, and highly supersoler iron abundance. 
We suggest that the spectrum of \hhfull is sculpted
more by absorption in a wind than by extreme relativistic effects
in strong gravity.
\end{abstract}


\section{Introduction}

\hhfull (hereafter \hh) is a Narrow Line Seyfert 1 (NLS1) galaxy
i.e. a low mass, high mass accretion rate (in terms of Eddington) AGN
\citep{Boroson2002}. It shows extreme dips in its X-ray lightcurve, during
which the spectra have a steep drop around 7~keV, associated with iron
K$\alpha$ \citep{Boller2002}. These spectra are generally fit with
ionized reflection, but the features are so strong and broad that this
interpretation requires extreme conditions.
The requirements are that the black hole spin is close to maximal,
that the incident radiation is strongly focussed onto the inner edge
of the disk whilst being suppressed in the direction of the observer,
and that iron is overabundant by a factor of 7--20 \citep{Fabian2004,
Fabian2009,Fabian2012,Zoghbi2010}. The first two features can be
explained together in a model where the dips are caused by an
extremely compact X-ray source on the spin axis of the black hole
approaching the event horizon (hereafter the lamppost model). The
resulting strong light bending focusses the intrinsic continuum away
from the observer (producing the drop in flux), whilst simultaneously strongly
illuminating the very inner disk \citep{Miniutti2004}.

However, the optical/UV continuum from \hh implies that the black hole
is accreting at a super-Eddington rate \citep{Done2016}. Hence the
inner disk is unlikely to be flat, as assumed in the lamppost
reflection models, and should launch a strong wind due to continuum
radiation pressure
\citep{Ohsuga2011,Jiang2014,Sadowski2016,Hashizume2015}.  Thus this
source, and other NLS1s with similarly very high mass accretion rates
and similarly extreme X-ray spectra
\citep[e.g. IRAS~13224-3809:][]{Ponti2010} may have spectral features
which are affected by absorption and/or emission/scattering in a
wind. Absorption has been persistently suggested as an alternative
explanation for the features seen around iron in the NLS1s, though the
spectra are complex and not easy to fit (Mkn~766:
\citealt{Miller2007,Turner2007}, MCG-6-30-15:
\citealt{Inoue2001,Gallo2004,Miller2008}, \hh:
\citealt{Boller2002,Miller2010,Mizumoto2014}).

The recent discovery of blueshifted ($v>10000$~km~s$^{-1}$ i.e. $0.03c$),
narrow absorption lines from very highly ionized material (mostly He-
and H-like Fe K$\alpha$) has focussed attention on winds from the
inner disk. These Ultra-Fast Outflows (UFOs) are seen from a variety
of nearby Seyfert galaxies \citep[see e.g. the compilation by][]{Tombesi2010}.
The fast velocities imply that the winds are launched from the
inner disk since winds typically have terminal velocities similar to the
escape velocity from their launch radius. However, the launch
mechanism is not well understood \citep[e.g.][]{Tombesi2012}. Standard broad
line Seyfert galaxies are well below their Eddington limit, so that they cannot
power winds from continuum radiation pressure alone. Moreover, they have strong
X-ray emission which ionizes the disk material above where it has
substantial UV opacity, suppressing a UV line driven disk wind \citep{Proga2000,Proga2004,Higginbottom2014}. Magnetic driving seems the only
remaining mechanism, but this depends strongly on the (unknown) magnetic field
configuration so is not yet predictive \citep{Proga2003}.

Despite this general lack of understanding of the origin of the winds,
the fastest, $v>0.2c$, and most powerful UFOs are seen in luminous
quasars such as \pds \citep{Reeves2009,Nardini2015} and \apm, a high
redshift quasar which is gravitationally lensed
\citep{Chartas2002}. These AGN are not standard broad line Seyfert
galaxies. Instead, since both quasars are around the Eddington limit,
they could power a continuum driven wind. Also, black holes in both
quasars are very high mass ($\sim 10^9-10^{10}$~\msun) so that their
accretion disk spectra should peak in the UV. \pds is also clearly
intrinsically X-ray weak \citep{Hagino2015}, and \apm may also be
similar (Hagino et al, in preparation). A more favorable set of
circumstances for UV line driving is hard to imagine. Nonetheless, the
extremely high ionization of the UFO in \pds means that the observed
wind material has no UV opacity, so UV line driving must take place
out of the line of sight if this mechanism is important
\citep{Hagino2015}.

Whatever the launch mechanism, \pds clearly has features at Fe
K$\alpha$ which are dominated by a wind from the inner disk rather
than extreme reflection from the inner disk
\citep{Reeves2003,Reeves2009,Nardini2015}. The best wind models so far
use Monte-Carlo techniques to track the complex, geometry and velocity
dependent, processes in the wind including absorption, emission,
continuum scattering and resonance line scattering
\citep{Sim2008,Sim2010,Hagino2015}. While these models have been used
to fit Mkn~766 \citep{Sim2008}, PG~1211+143 \citep{Sim2010}, \pds
\citep{Hagino2015}, and six other `bare' Seyfert 1 nuclei
\citep{Tatum2012}, they have not yet been applied to \hh, the object
with the strongest and broadest iron features, and the one where the
reflection models require the most extreme conditions
\citep{Fabian2009,Fabian2012}. Previous work showed that simple wind
models, where the iron features were described using a single P Cygni
profile, can fit the deep drop at $\sim 7$~keV seen in one observation
of this object \citep{Done2007}, but here we use the full Monte-Carlo
wind code, \monaco, of \cite{Hagino2015} to see if this can adequately
fit all the multiple observations of \hh. We show that inner disk wind
models can indeed give a good overall description of the 2--10~keV
spectra, and that extrapolating these to higher energies can also
match the \nustar data from this source.

\section{Observations of \hh and comparison to the archetypical
  wind source \pds}

\begin{table}
\caption{\xmm and \suzaku observations of \hh}
\centering
\begin{tabular}{lllr}
\hline\hline
Name & Obs ID & Start Date & Net exposure (ks)\footnotemark[1]\\
\hline
\multicolumn{4}{c}{\xmm}\\
\hline
Obs1 & 0110890201 & 2000-10-21 & 37.8 \\
Obs2 & 0148010301 & 2002-10-13 & 68.1 \\
Obs3 & 0506200301 & 2007-05-14 & 35.8 \\
Obs4 & 0506200201 & 2007-05-16 & 26.9 \\
Obs5 & 0506200501 & 2007-06-20 & 32.6 \\
Obs6 & 0506200401 & 2007-07-06 & 14.7 \\
Obs7 & 0511580101 & 2008-01-29 & 99.6 \\
Obs8 & 0511580201 & 2008-01-31 & 66.4 \\
Obs9 & 0511580301 & 2008-02-02 & 59.8 \\
Obs10 & 0511580401 & 2008-02-04 & 66.6 \\
Obs11 & 0653510301 & 2010-09-13 & 103.7 \\
Obs12 & 0653510401 & 2010-09-15 & 102.1 \\
Obs13 & 0653510501 & 2010-09-17 & 95.8 \\
Obs14 & 0653510601 & 2010-09-19 & 97.7 \\
Obs15 & 0554710801 & 2011-01-12 & 64.5 \\
\hline
\multicolumn{4}{c}{\suzaku}\\
\hline
SuzakuObs & 700008010 & 2005-12-03 & 97.9/100.4/97.2/97.8\\
\hline
\end{tabular}
\begin{flushleft}
\footnotesize
$^1$ Net exposure time of pn for \xmm and XIS0/XIS1/XIS2/XIS3 for \suzaku, respectively.
\end{flushleft}
\label{tab:allobs_1h}
\end{table}%

\begin{figure}
\centering
\includegraphics[width=\hsize]{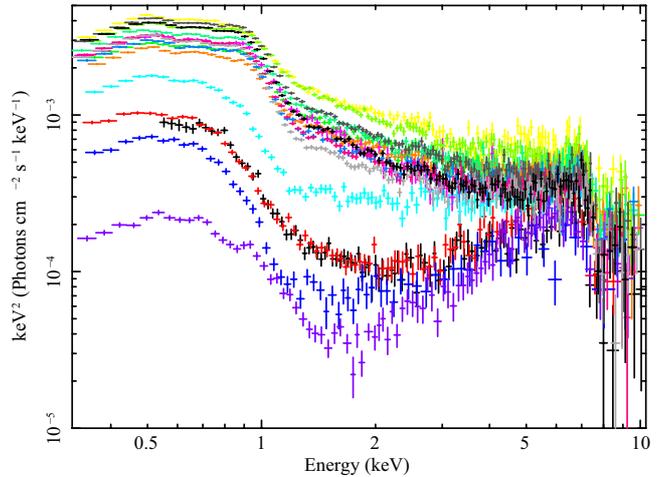}
\caption{All 2--10~keV spectra observed by \xmm/pn and \suzaku/FI detectors, unfolded against a $\Gamma=2$ power law.}
\label{fig:allspec_1h}
\end{figure}

\hh was observed by \suzaku and \xmm for many times as listed in
Tab.~\ref{tab:allobs_1h}.
We reduce both \xmm pn and \suzaku XIS data with standard screening conditions:
${\rm PATTERN}\leq4$ events for \xmm/pn data and grade 0, 2, 3, 4 and 6 events for \suzaku/XIS data were used.
Bad time intervals were also excluded.
\xmm data in time intervals when background rates of ${\rm PATTERN}=0$ events at energy $>10$~keV exceeded 0.4~counts~s$^{-1}$ were removed. \suzaku data within 436~s of passage through the South Atlantic Anomaly (SAA), and within an Earth elevation angle (ELV) $<5^\circ$ and Earth day-time elevation angles (DYE\_ELV) $<20^\circ$ were excluded.
Spectra were extracted from circular regions of 64$\arcsec$ and 2.$\arcmin$9 diameter for \xmm and \suzaku, respectively. Background spectra of \xmm data were extracted from circular regions of the same diameter in the same chip as the source regions, while background spectra of \suzaku data were extracted from annular region from 7.$\arcmin$0 to 14.$\arcmin$0 diameter.
The observed data show a large
variability in continuum spectra as seen in
Fig.~\ref{fig:allspec_1h}. There is an obvious variability in the
2--6~keV continuum shape and in the strength of the iron K features
around 7~keV.

\begin{figure}
\begin{center}
\includegraphics[width=\hsize]{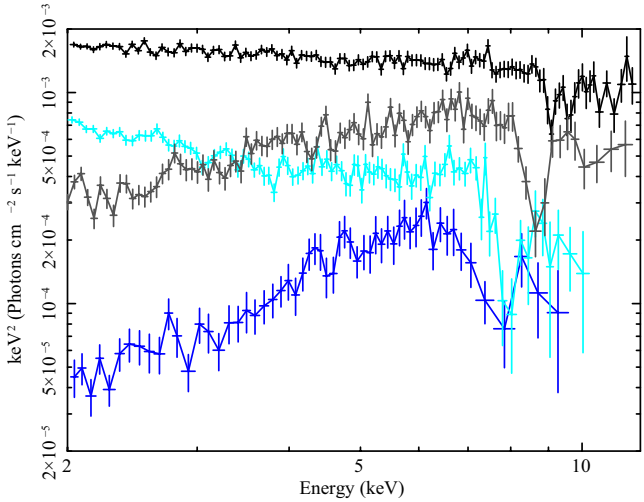}
\caption{Comparison between the brightest/faintest spectra of
\pds (black/grey: ObsID 707035030/701056010) and the
brightest/faintest spectra of \hh (cyan/blue: Obs12/Obs15). The
2--10~keV spectra show a similar range in 2-5~keV spectral slope, and
similarly deep features at Fe K$\alpha$, though these start at a lower
energy in \hh, and are broader than in \pds.}
\label{fig:compare}
\end{center}
\end{figure}

Figure~\ref{fig:compare} shows brightest and faintest 2--12~keV (rest
frame) spectra seen in \pds, the archetypal inner disk wind source,
(black/grey: ObsID 707035030/701056010), together with the
corresponding brightest/faintest observations of \hh (cyan/blue:
Obs12/Obs15).  The spectra are similarly harder when fainter, and show
a similarly deep drop at iron K. In \pds, these features are all generally
associated with absorption, with the deep drop at iron K interpreted
as Fe K$\alpha$ absorption lines from highly ionized material in the
wind, while the variability at lower energies can be described by
complex absorption from lower ionization material \citep{Reeves2003,Reeves2009,Nardini2015,Hagino2015}. Conversely, the spectral
features in \hh are generally associated with extreme relativistic
reflection, especially in the hardest/dimmest spectra \citep{Fabian2009,Fabian2012}. Nonetheless, the 2--10~keV spectra of 
 \hh and \pds are quite similar, both in range of continuum shapes
 and in the shape of the drop around Fe K$\alpha$.

However, there are also some more subtle differences. The sharp drop in \hh
is at $\sim7.0\textrm{--}7.5$~keV, whereas it is at $8.5\textrm{--}9.0$~keV in \pds, and
the recovery of the continuum after the absorption lines at higher
energies is more evident in \pds than in \hh. Thus if these
features are from a wind in \hh, it shares many similarities with
the wind in \pds but cannot have exactly the same
parameters.

\section{The \monaco simulations for \pds}

The wind in \pds was modeled by using the \monaco Monte-Carlo
code. This wind model and the calculation scheme are fully described
in \cite{Hagino2015}, and the general framework design of the code is
described in \cite{Odaka2011}.  To summarize its main features, it
follows \cite{Sim2010} and assumes a biconical wind geometry, where
the wind is launched from radii $R_{\rm min}$ to $R_{\rm max}$ along
streamlines which diverge from a focal point at a distance $d$ below
the black hole. The radial velocity of material along each streamline
of length $l$ from its launch point on the disk is
\begin{eqnarray}
v_{\rm r}(l)=v_0+(v_\infty-v_0)\left(1-\frac{R_{\rm min}}{R_{\rm min}+l}\right)^\beta,\label{eq:velocity}
\end{eqnarray}
while
the azimuthal velocity is assumed to be Keplarian at the launch point,
and then conserves angular momentum as the wind expands. We assume $v_0$ is
negligible. There can also be a turbulent velocity, $v_{\rm t}$.

The total wind mass loss rate $\dot{M}_{\rm wind}$ is given as
\begin{eqnarray}
\dot{M}_{\rm wind}=1.23m_{\rm p} n(r) v_{\rm r}(r) 4\pi D^2 \frac{\Omega}{4\pi},\label{eq:density}
\end{eqnarray}
which sets the density $n(r)$ as a function of radius in
the wind. Here, $1.23m_{\rm p}$, $D$ and $\Omega$ are an ion mass, the distance from the focal point and a solid angle of the wind, respectively. The bicone is split into 100 radial shells, and the ion
populations are calculated in each shell by using the {\sc xstar}
photoionization code on the assumption that the central source is a power law spectrum with photon index $\Gamma$ and ionizing luminosity $L_{\rm x}$ (adjusted for special
relativistic dimming from the radial outflow velocity). 
The resulting H- and He-like ion densities of each element are put in the Monte-Carlo radiation transfer simulator \monaco, which uses the {\sc Geant4} toolkit library \citep{Allison2006} for photon tracking,
but extended to include a full treatment
of photon processes related to H- and He-like ions \citep{Watanabe2006}.

Thus there are 9 major free parameters. $R_{\rm min}$, $R_{\rm max}$ and $d$
determine the geometry. $v_\infty$, $v_{\rm t}$ and $\beta$ define the
radial velocity structure. These geometry and velocity structure together with $\dot{M}_{\rm wind}$
determine the density structure, which then sets the ionization state
given $L_{\rm x}$ and $\Gamma$.

This wind model is self-similar in ionization structure and column density for
systems at different mass but the same Eddington ratio $\dot{m}=L_{\rm bol}/L_{\rm Edd}$
i.e. the same geometry, velocity structure and spectral index. The
ionization state $\xi=L_{\rm x}/nD^2$ can be written as
\begin{eqnarray}
\xi = 1.23 m_{\rm p} v_{\rm r} \Omega L_{\rm x}/\dot{M}_{\rm wind},
\end{eqnarray}
where equation~\ref{eq:density} is used.
Since the ionizing luminosity can be written as $L_{\rm x}=f_{\rm x}\dot{m}\eta \dot{M}_{\rm Edd} c^2$
where $f_{\rm x}=L_{\rm x}/L_{\rm bol}$,
then the ionization parameter reads
\begin{eqnarray}
\xi=1.23 m_{\rm p} v_{\rm r}\Omega f_{\rm x}\dot{m} \eta c^2/(\dot{M}_{\rm wind}/\dot{M}_{\rm Edd}).
\end{eqnarray}
Thus the
ionization state of the wind is determined by
$\dot{M}_{\rm wind}/\dot{M}_{\rm Edd}$, and not by black hole mass directly.

The observed spectral properties strongly depend on the inclination angle.
The energy at which the line absorption is seen depends on both the
terminal wind speed and the inclination angle at which we see the
wind, whereas the width of the absorption line depends on the spread of
velocities along the line of sight. This spread in turn depends on the
terminal velocity of the wind, how much of the acceleration region is
along the line of sight, and the projected angle between the wind
streamlines and the line of sight. Along the bicone, a fast wind
acceleration law means that most of the wind is at its terminal
velocity. Thus the line width is fairly small and the blueshift indicates
the true wind velocity. At higher inclination angles, the line of
sight cuts across the acceleration region so the absorption line is wider and
the total blueshift is not so large since the line of sight includes much lower velocity material.
Thus the same wind seen at different line of sight can have
very different properties in terms of the measured width and blueshift
of the absorption line. 

\begin{figure*}
\begin{center}
\includegraphics[width=\hsize]{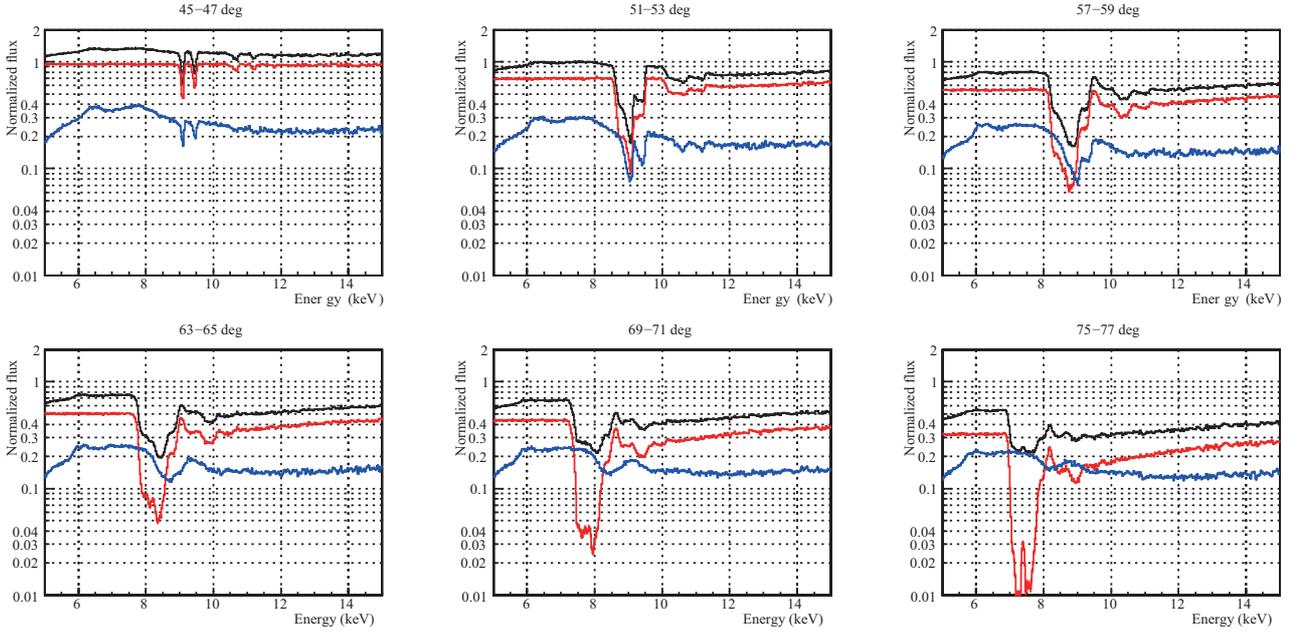}
\caption{\monaco wind model for \pds seen at different
inclination angles. The direct and reprocessed components are plotted in red and blue, respectively. The sums of these components are plotted in black.
Angles from $45-47^\circ$ intercept only the top
of the bicone. Almost all the wind is at the terminal velocity of
$0.3c$ so the lines are strongly blueshifted, but fairly
narrow. Larger angles cut through more of the wind, so the lines are
deeper, and intersect progressively more of the acceleration zone, so
the lines are broader and blend into each other. Angles larger than
$56.6^\circ$ now cut across the bicone, so the lines are less
blueshifted, but include even more of the acceleration zone, so the
lines are broader, and now the wind is close to optically thick so the
edge is also important. Thus the line absorption shifts from two
obviously narrow lines which are strongly blueshifted, to a single,
very broad absorption line which is not strongly blueshifted.
}
\label{fig:pdssim}
\end{center}
\end{figure*}

We illustrate this inclination dependence in Fig.~\ref{fig:pdssim} by showing the \monaco wind
model for \pds at a larger range of inclination angles than in
\cite{Hagino2015}. This wind model has $R_{\rm min}=d=20R_{\rm g}$ and $R_{\rm max}=1.5R_{\rm min}$
so that the wind fills a bicone between $\theta_{\rm min}=45^\circ$ to
$\theta_{\rm max}=56^\circ.3$, i.e. the wind fills a solid angle $\Omega/4\pi=0.15$. We fix
the velocity law at $\beta=1$ and assume $v_0=v_{\rm t}=10,000$~km/s.
The terminal velocity is set at $v_\infty=-0.3c$, which is consistent with the escape
velocity from $R_{\rm min}$. The mass outflow rate is $\dot{M}_{\rm wind}/\dot{M}_{\rm Edd}=0.3$, and the ionizing spectrum with $L_{\rm x}/L_{\rm Edd}=10^{44}/10^{47}$ and $\Gamma=2.5$ is assumed. It is clear that inclination
angles larger than $\theta_{\rm max}$ gives a line which is less
blueshifted, but broader, as required to fit \hh. The width is
accentuated because the larger column density of the wind means that the
ionized iron edge is also important in absorbing the spectrum blueward
of the absorption line. These high inclination \monaco model spectra 
for \pds give iron K features which are quite similar to those
required for \hh.

\section{Fitting the range of 2--10~keV data from \hh} \label{sec:fitting}
\subsection{Tailoring the wind model to \hh}

\begin{figure*}
\begin{center}
\includegraphics[width=0.49\hsize]{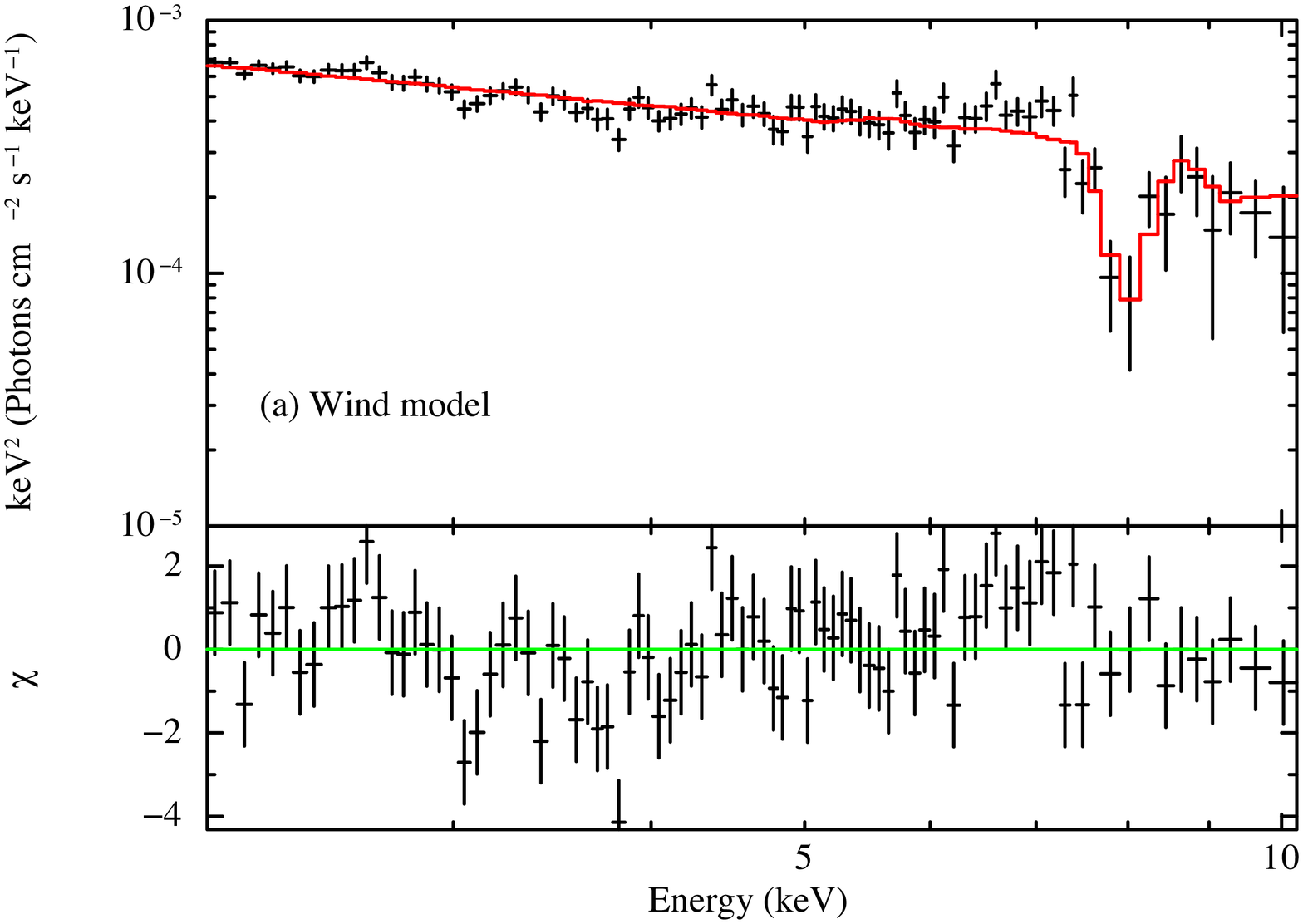}
\includegraphics[width=0.49\hsize]{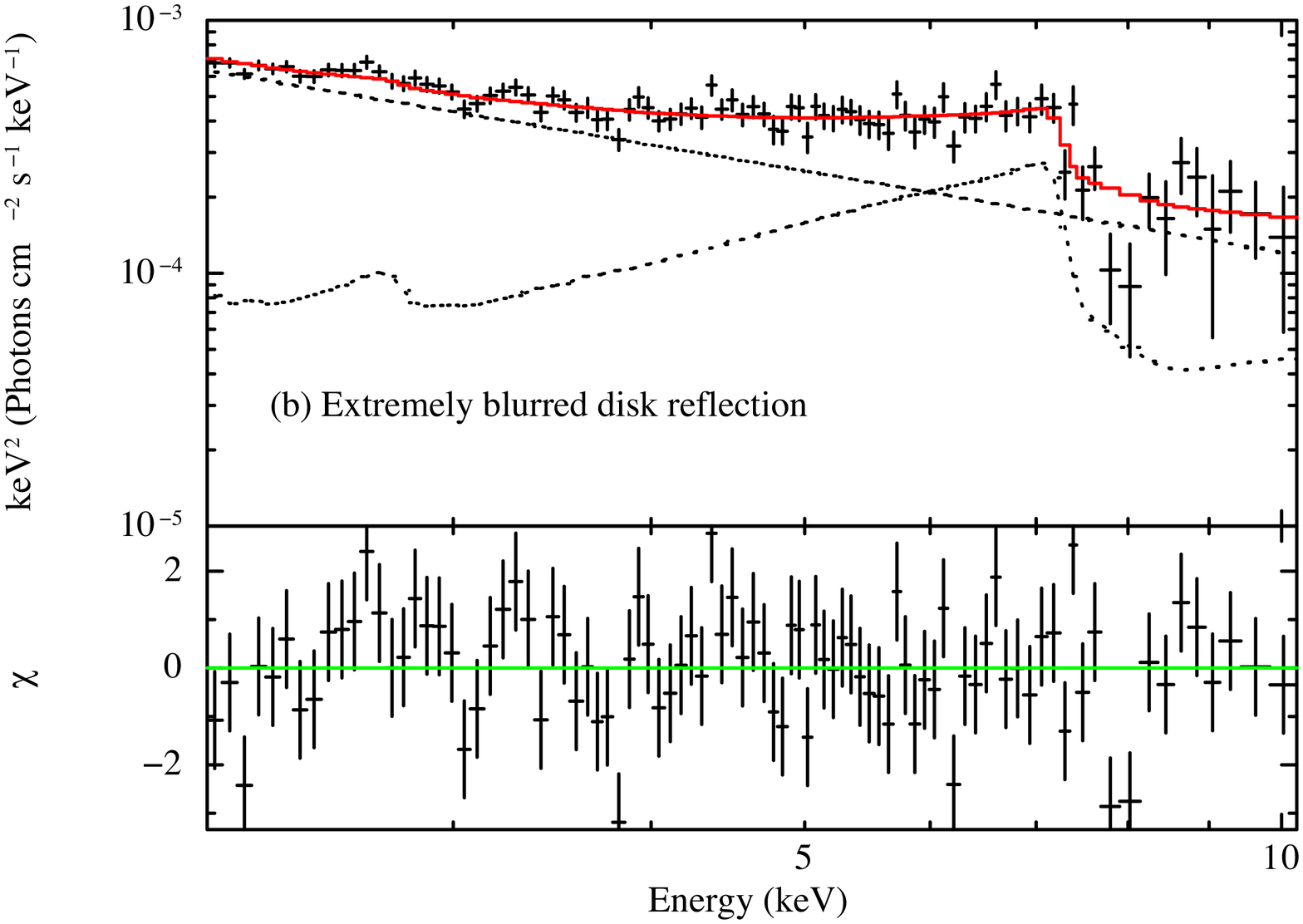}
\includegraphics[width=0.49\hsize]{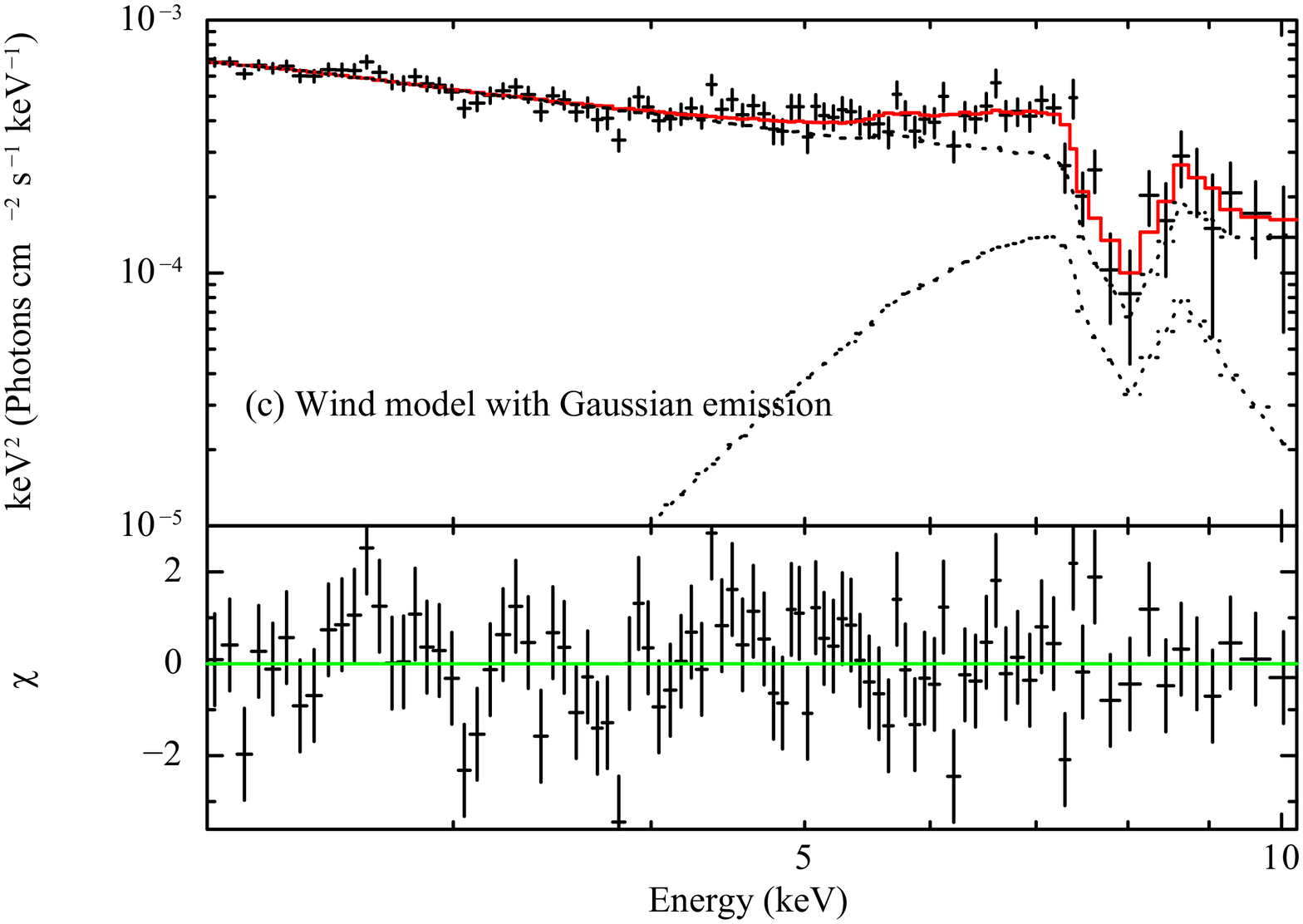}
\includegraphics[width=0.49\hsize]{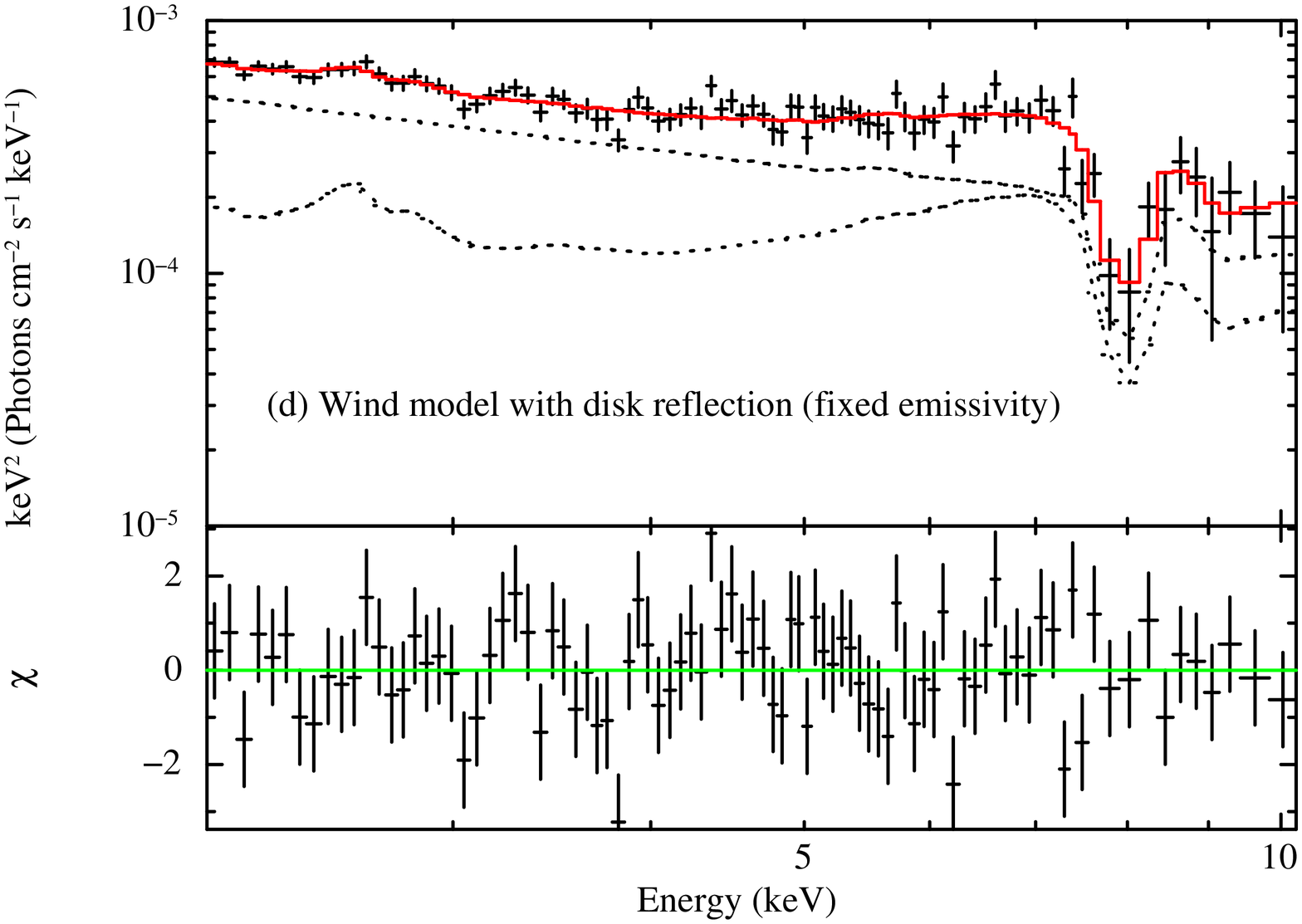}
\caption{\xmm/pn spectra of Obs12 fitted by (a) {\sc wind*powerlaw} (top-left),
(b) {\sc powerlaw+kdblur*reflionx} (top-right), (c) {\sc wind(powerlaw+zgaus)}
(bottom-left) and (d) {\sc wind(powerlaw+kdblur*reflionx)} with emissivity fixed at 3
(bottom-right).
The spectra and models are plotted in black and red, respectively.
The lower panels show the residuals in units of $\chi$.
}
\label{fig:obs12}
\end{center}
\end{figure*}

We first see whether the \monaco model calculations for the archetypal wind
source \pds can fit the features seen in \hh by simply viewing
the wind at a larger inclination angle. We fit to Obs12, which is
the steepest spectrum seen from \hh. We assume that this
steepest spectrum has negligible absorption from lower ionization species.
We use the \monaco results from \pds, as coded into a multiplicative
model by \cite{Hagino2015}, on a power law continuum. Free parameters
are the inclination and the redshift. By allowing the redshift to be free,
we are able to fit for a slightly different wind velocity. This fitting gives
${\chi_\nu}^2=133/87$ for a power law index of $2.65\pm 0.05$ (see
Fig.~\ref{fig:obs12}a and Table~\ref{tab:obs12}).

We compare this model with the standard extreme reflection
interpretation. We use the ionized reflection {\sc atable} models {\sc
reflionx} \citep{Ross2005}, and convolve these with {\sc kdblur},
allowing the emissivity index, inner disk radius, inclination angle,
iron abundance and ionization parameter to be free parameters.
This gives ${\chi_\nu}^2=115/83$  for $\Gamma=3.07\pm 0.15$, a
somewhat better fit, but with more free parameters (Fig.~\ref{fig:obs12}b). Inspection of the
residuals shows that the wind model underpredicts the emission line
flux, whereas the reflection model underpredicts the extent of the
drop at $\sim 7.5$~keV. This suggests that the drop is better matched
by the wind, but there is more line emission than predicted in the
\pds models. 

This discrepancy could be produced by the wind itself. A wider angle wind will intercept
more of the source luminosity, and have stronger emission/reflection/scattering features at iron K
for a given absorption line strength. We will investigate such wide angle winds in a subsequent paper.
Alternatively (or additionally), there can be lower ionization species in the wind which also add to the 
line emission. We approximate both of these physical pictures by adding a 
phenomenological broad gaussian line to the wind model, and find a better fit than either 
reflection or 
the \pds wind alone (${\chi_\nu}^2=107/84$: Fig.~\ref{fig:obs12}c). 

Another possible origin for the excess iron line emission is that there is 
residual reflection from the underlying disk. We add a phenomenological
blurred reflection component as above, allowing the amount of disk reflection,
emissivity index, inner disk radius, inclination angle, iron abundance and
ionization parameter to be free. This is a better fit than the gaussian line
($\chi^2_\nu=92.0/82$), mostly because the reflection model also includes
hydrogen-like sulphur line emission at 2.8~keV which is clearly present
in the data. However, the best fit reflection parameters are
now more extreme than those for the fit without a wind, with strongly centrally
peaked emissivity and fairly small inner radius (Table~\ref{tab:obs12}).

\begin{figure*}
\begin{center}
\includegraphics[width=0.49\hsize]{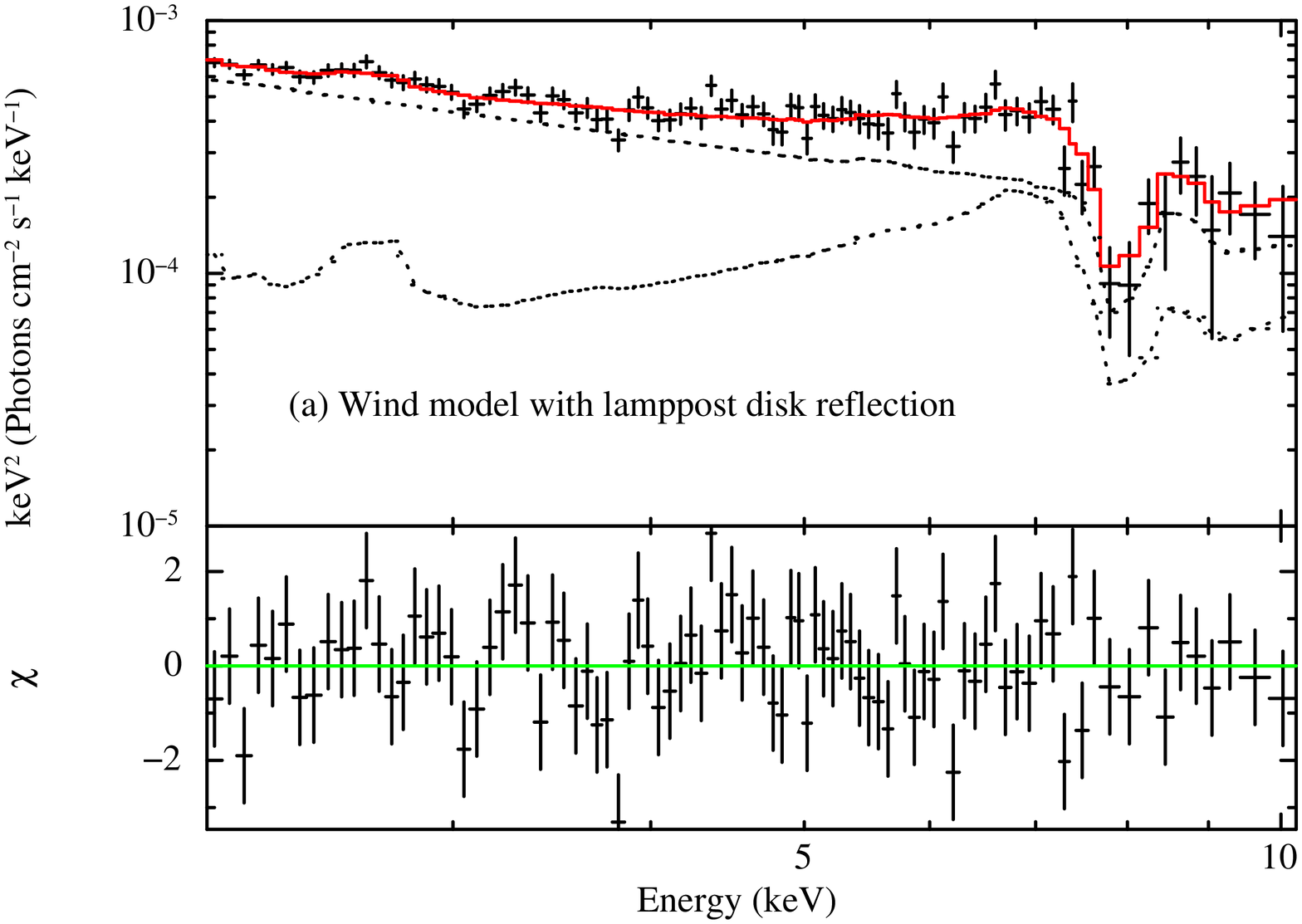}
\includegraphics[width=0.49\hsize]{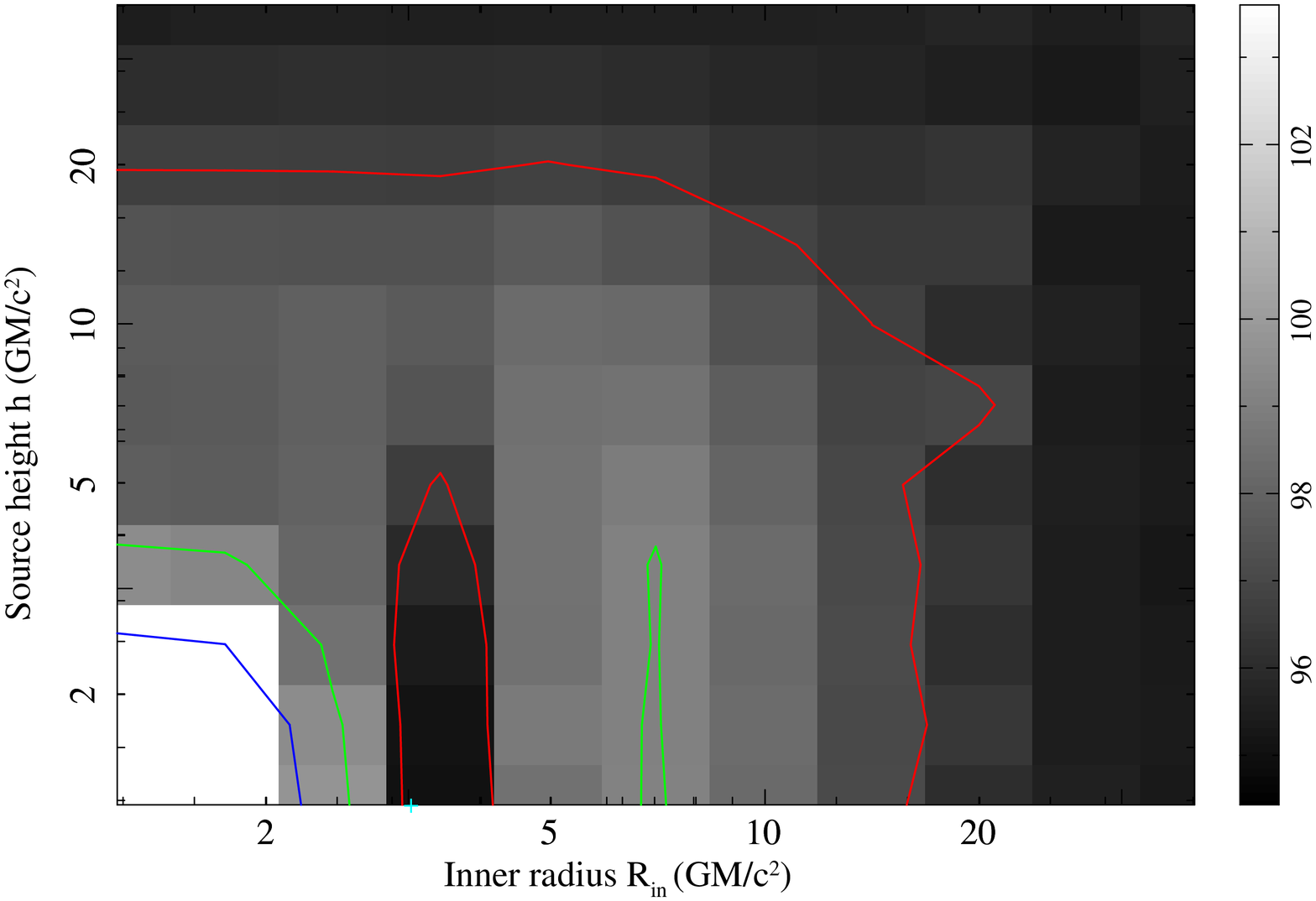}
\caption{
\xmm/pn spectra of Obs12 fitted by the wind model with lamppost disk reflection
({\sc wind(powerlaw+relconv\_lp*reflionx)}, left panel) and contour plot of
the inner disk radius and the height of the X-ray compact source in
the lamppost geometry (right panel).
The red, green and blue lines in the contour plot correspond to confidence levels of 68, 90 and 99 percent, respectively. The color map shows the value of $\chi^2$ for each parameter grid.
The cyan cross is plotted at the best fit, where $h\sim1$~$r_{\rm g}$ and $R_{\rm in}\sim3$~$r_{\rm g}$. The model spectrum shown in the left panel corresponds to this extreme solution.
}
\label{fig:lamppost}
\end{center}
\end{figure*}

Such strongly centrally peaked emissivity is characteristic of the lamppost model,
but this produces an illumination emissivity profile which is more complex than
a single power law. Hence we replace {\sc kdblur} with the fully relativistic
lamppost model for the emissivity \citep[{\sc relconv\_lp};][]{Dauser2013}.
Fixing the black hole spin to maximal and allowing the inner radius
to be free (as in {\sc kdblur}) gives a slightly worse fit, with $\chi^2_\nu=94.4/82$
for similarly extreme parameters (inner radius of $\sim3$~$r_{\rm g}$ and
height of $\lesssim2$~$r_{\rm g}$: Fig.~\ref{fig:lamppost}a). 
However, this fit is very poorly constrained as shown in Fig.~\ref{fig:lamppost}b.
Only the most extreme solution at small $R_{\rm in}$ and small $h$ can be excluded with
more than 99\% confidence level, while the best fit has very similar fit statistic
for much less extreme fits ($R_{\rm in}\sim 30$~$r_{\rm g}$ for height which is
unconstrained). We illustrate a statistically equivalent less extreme fit in Fig.~\ref{fig:obs12}d by using {\sc kdblur} with fixed emissivity of 3 and find a much larger inner radius of $\sim 36$~$r_{\rm g}$ ($\chi^2_\nu=95.1/83$: Table~\ref{tab:obs12}).
This result shows directly that with
wind absorption, the lamppost reflection is not required to be extreme in either
relativistic smearing nor in iron abundance.

Regardless of how the additional emission component is modeled, 
our wind model developed for the highly blueshifted absorption lines in \pds
is better able to produce the observed sharp drop at $\sim 7$~keV in \hh than relativistic reflection alone.

\begin{table*}
\caption{Fitting parameters for Obs12 with \monaco wind models and the extremely blurred reflection model.}
\begin{center}
\begin{tabular}{llccccc}
\hline\hline
& & \multirow{3}{*}{(a) \monaco wind} & (b) {\sc kdblur} & (c) \monaco wind & \multicolumn{2}{c}{(d) \monaco wind}\\
& &  & {\sc *reflionx} &  + {\sc zgaus} &  \multicolumn{2}{c}{+ {\sc kdblur*reflionx}}\\
& &  &  & & Emissivity:free &  Emissivity:3\\
\hline
\monaco wind & Velocity ($c$) & $0.19_{-0.02}^{+0.02}$ & --- & $>0.17$ & $0.18_{-0.02}^{+0.02}$ & $0.18_{-0.02}^{+0.04}$\\
& $\theta_{\rm incl}$ ($^\circ$) & $52.6_{-2.0}^{+2.7}$ & --- & $62.4_{-10.9}^{+2.0}$ & $51.4_{-2.1}^{+2.0}$ & $52.4_{-3.1}^{+3.6}$\\
Powerlaw & $\Gamma$  & $2.65_{-0.04}^{+0.05}$ & $3.07_{-0.12}^{+0.18}$ & $2.83_{-0.10}^{+0.09}$ & $2.89_{-0.10}^{+0.15}$ & $2.81_{-0.19}^{+0.19}$\\
\hline
{\sc kdblur} & Index & --- & $3.57_{-0.35}^{+0.66}$ & ---  & $>4.39$ & $3.0$\footnotemark[1]\\
& $R_{\rm in}$ ($r_{\rm g}$) & --- & $<10.85$ & --- & $3.3_{-0.6}^{+0.6}$ & $36_{-28}^{+157}$\\
& $\theta_{\rm incl}$ ($^\circ$) & --- & $48.90_{-4.16}^{+3.58}$ & --- & tied to wind & tied to wind\\
\hline
{\sc reflionx} & Fe abundance & --- & $>6.75$ & --- & $3.01_{-1.68}^{+4.79}$ & $1.58_{-1.36}^{+5.72}$\\
& $\xi$ & --- & $99.6_{-44.8}^{+118.1}$ & --- & $118_{-71}^{+187}$ & $699_{-650}^{+534}$\\
\hline
Gaussian & Line energy & --- & --- & $6.68_{-0.62}^{+0.63}$ & --- & ---\\
& Sigma & --- & --- & $>1.18$ & --- & ---\\
\hline
Fit statistics & $\chi^2$/dof & $133.4/87$ & $115.5/83$ & $107.4/84$ & $92.0/82$ & $95.1/83$\\
& Null prob. & ${1.0\times10^{-3}}$ & ${1.1\times10^{-2}}$ & ${4.4\times10^{-2}}$ & $0.21$ & $0.17$\\
\hline
\end{tabular}
\end{center}
\begin{flushleft}
\footnotesize
$^1$ Parameters are fixed.
\end{flushleft}
\label{tab:obs12}
\end{table*}

\subsection{Application to \xmm and \suzaku data}

We re-run \monaco with the parameters around the values obtained in
the previous section but now put in explicitly the mass of $5\times
10^6$~\msun, an observed X-ray spectrum with $\Gamma=2.6$ and $L_{\rm
  x}=2\times 10^{42}$~erg~s$^{-1}$, and reduce the wind velocity to
$0.2c$ but keep the same velocity law.

The result from the previous section suggests that there is rather
more line emission compared to absorption than is predicted for the
\pds wind geometry. However, a larger solid angle of the wind would
necessitate more resolution in $\theta$ and a change in the way of
calculating the ionization state. Hence we keep the same wind geometry
for the present, and run simulations for $\dot{M}_{\rm
  wind}/\dot{M}_{\rm Edd}=0.13$, $0.20$, $0.26$. Refitting these to
Obs12 gives $\chi^2=139.7, 132$ and $133$ for $89$ degrees of freedom,
respectively. Hence we pick $\dot{M}_{\rm wind}/\dot{M}_{\rm
  Edd}=0.2$, and apply this model to all the observations of \hh.
Similarly to \pds, we assume that the intrinsic power law stayed
constant in spectral index, and that the major change in spectral
hardness is produced by a changing absorption from lower ionization species.
To describe this absorption, we used a very simple model
of partially ionized absorption which can partially cover the source
({\sc zxipcf} in {\sc xspec}). This gives an adequate fit to all the
\pds spectra, including the hardest (black line in
Fig.~\ref{fig:compare}) which bears some similarities to the hardest
spectra in \hh, which are generally interpreted as reflection
dominated \citep{Fabian2012}.

\begin{figure*}
\begin{center}
\includegraphics[width=0.31\hsize]{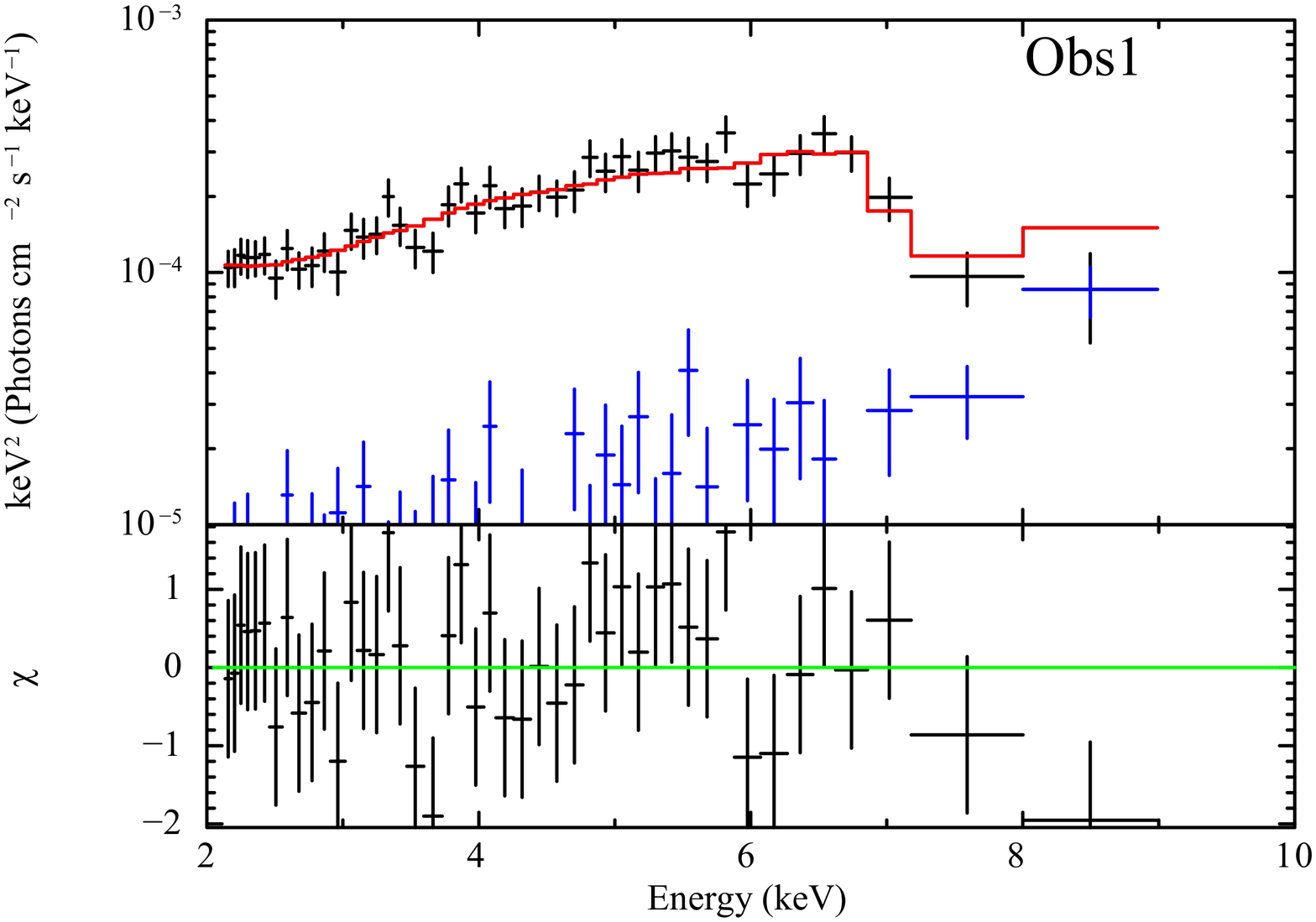}
\includegraphics[width=0.31\hsize]{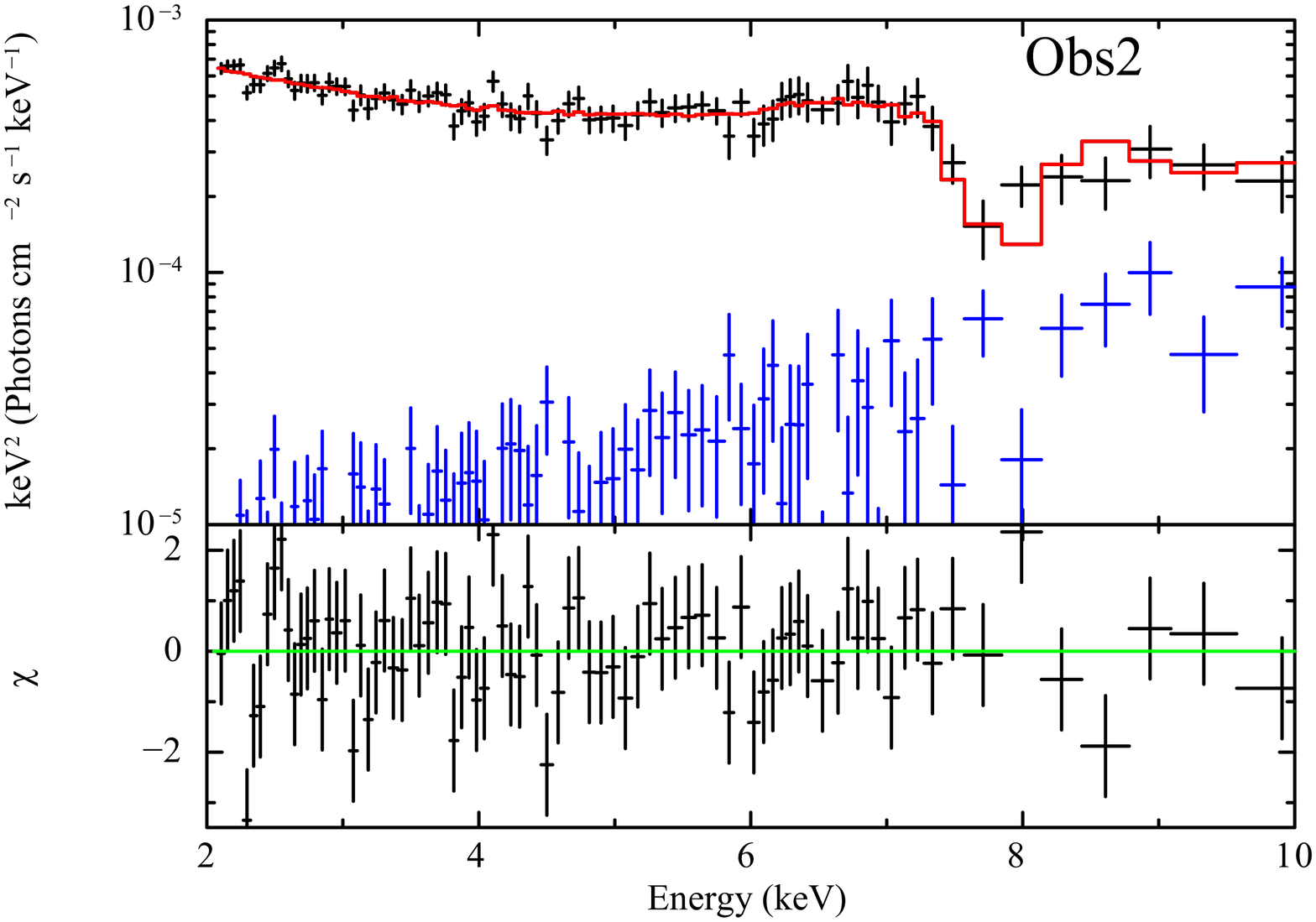}
\includegraphics[width=0.31\hsize]{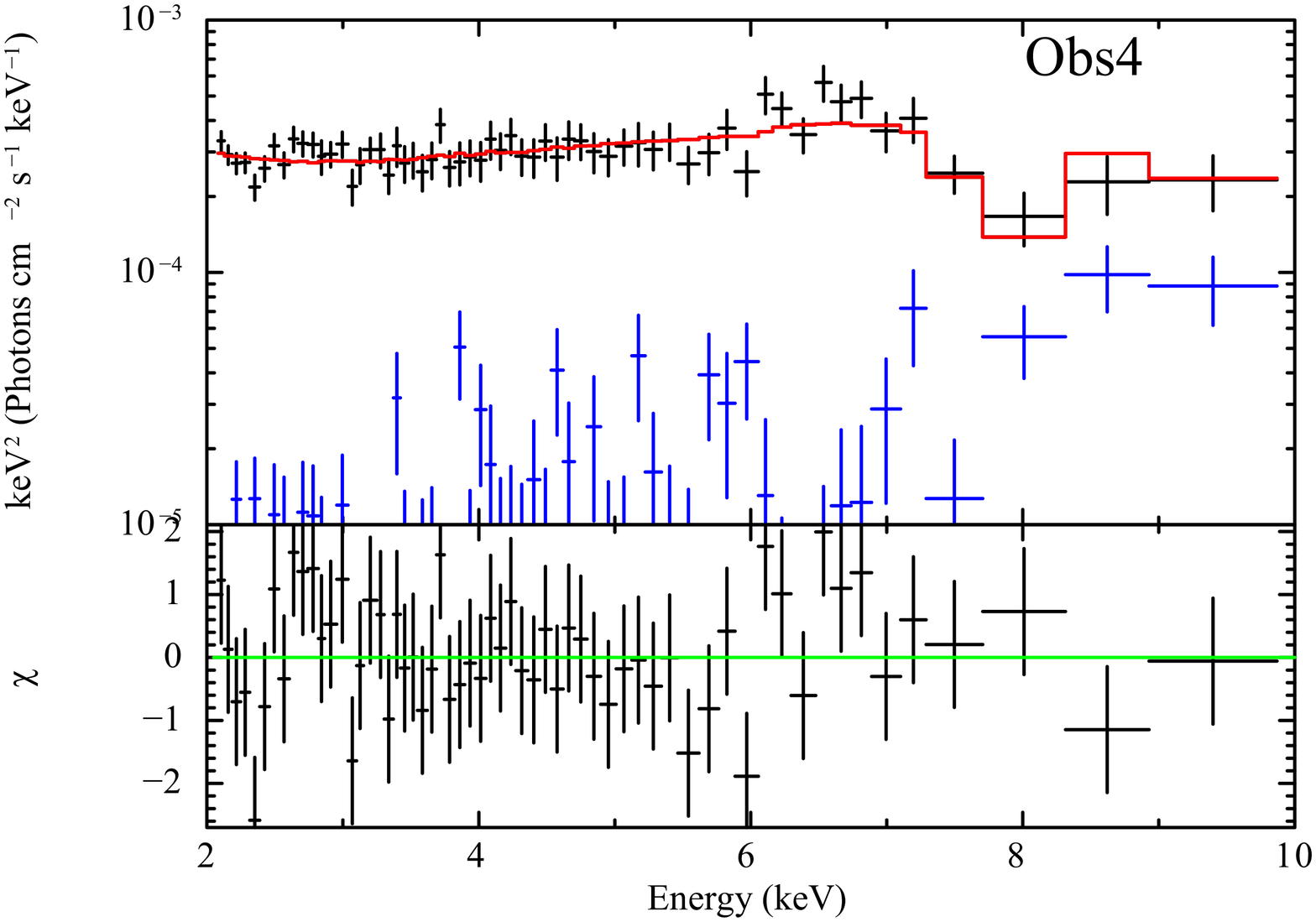}

\includegraphics[width=0.31\hsize]{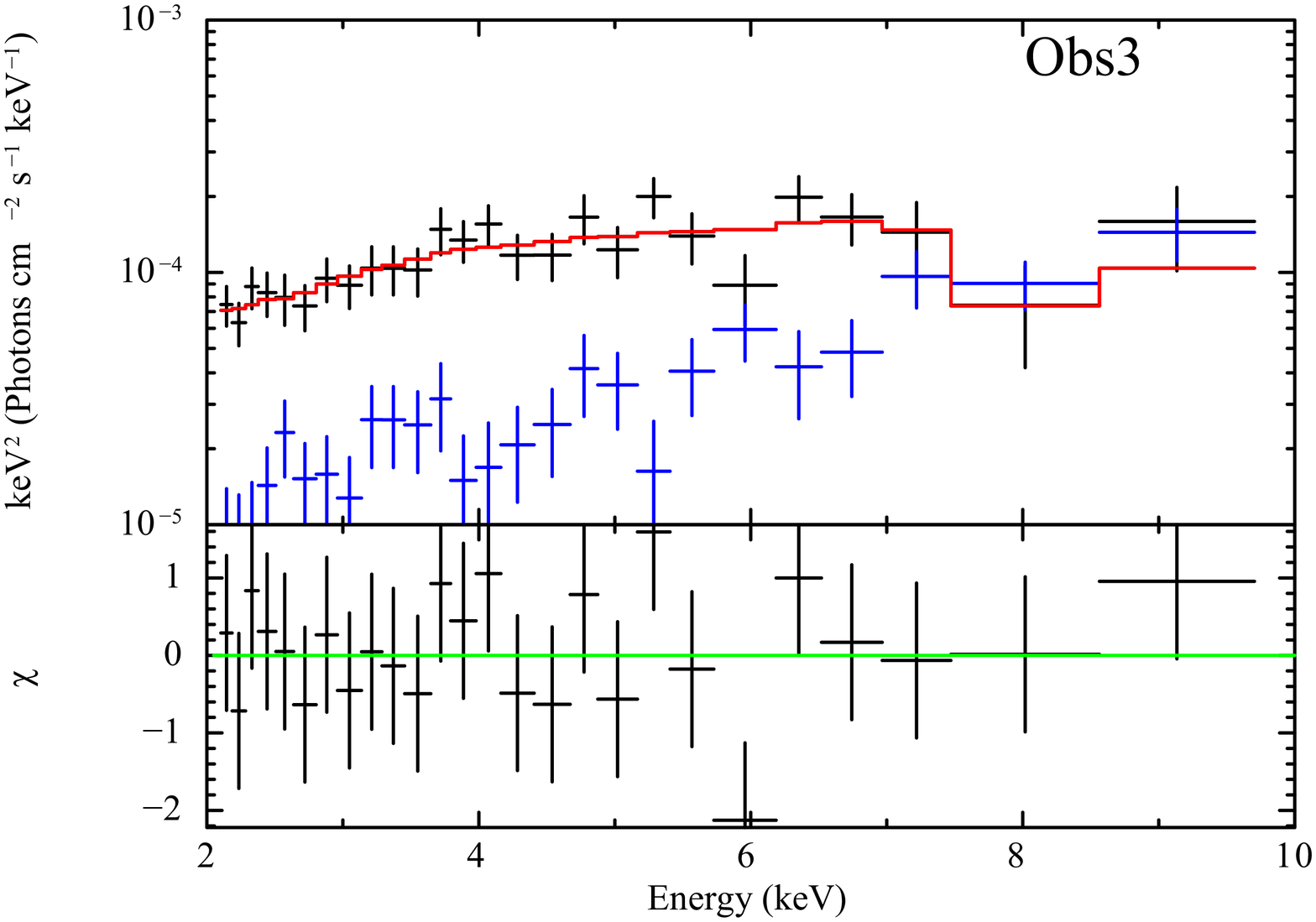}
\includegraphics[width=0.31\hsize]{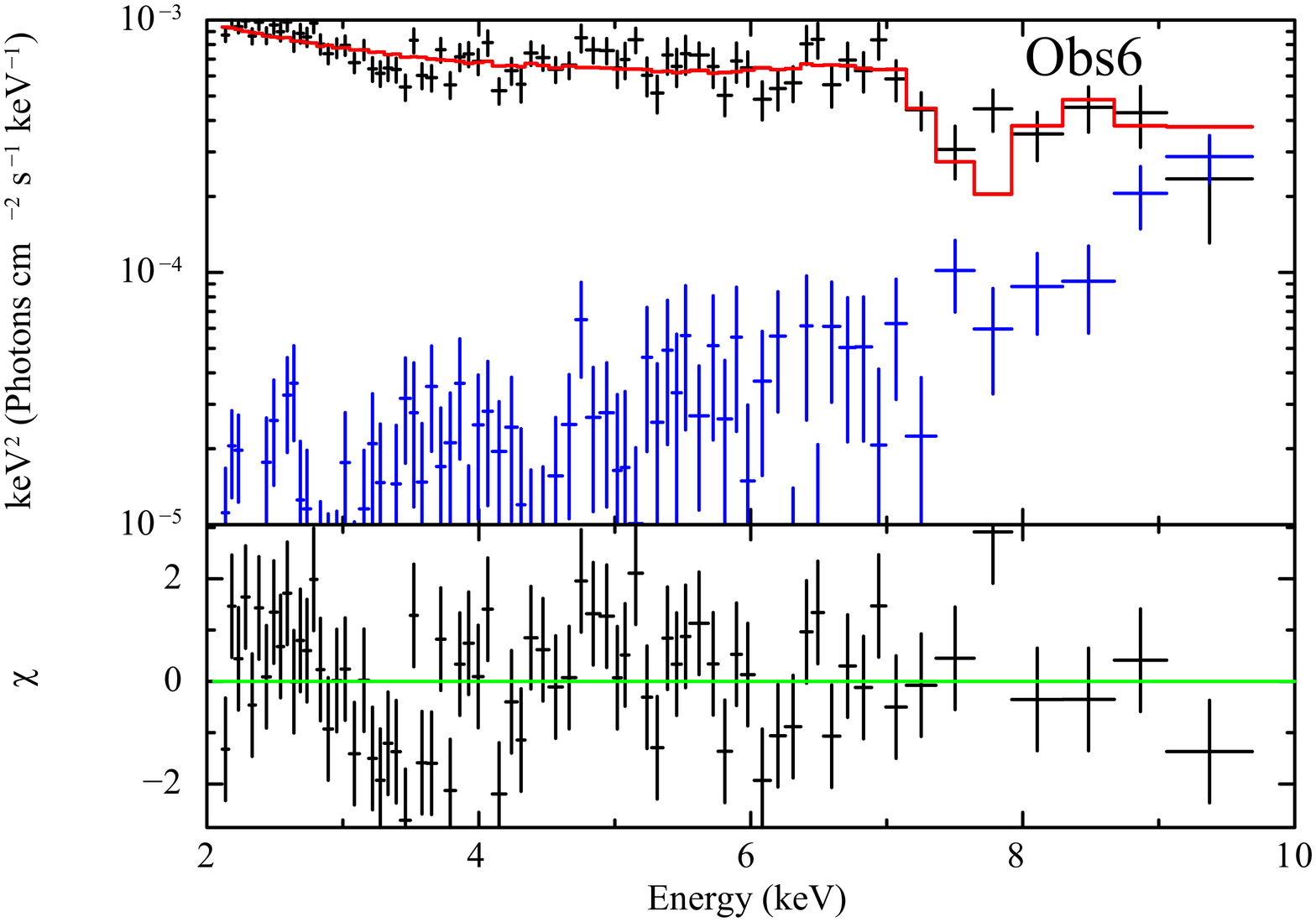}%5
\includegraphics[width=0.31\hsize]{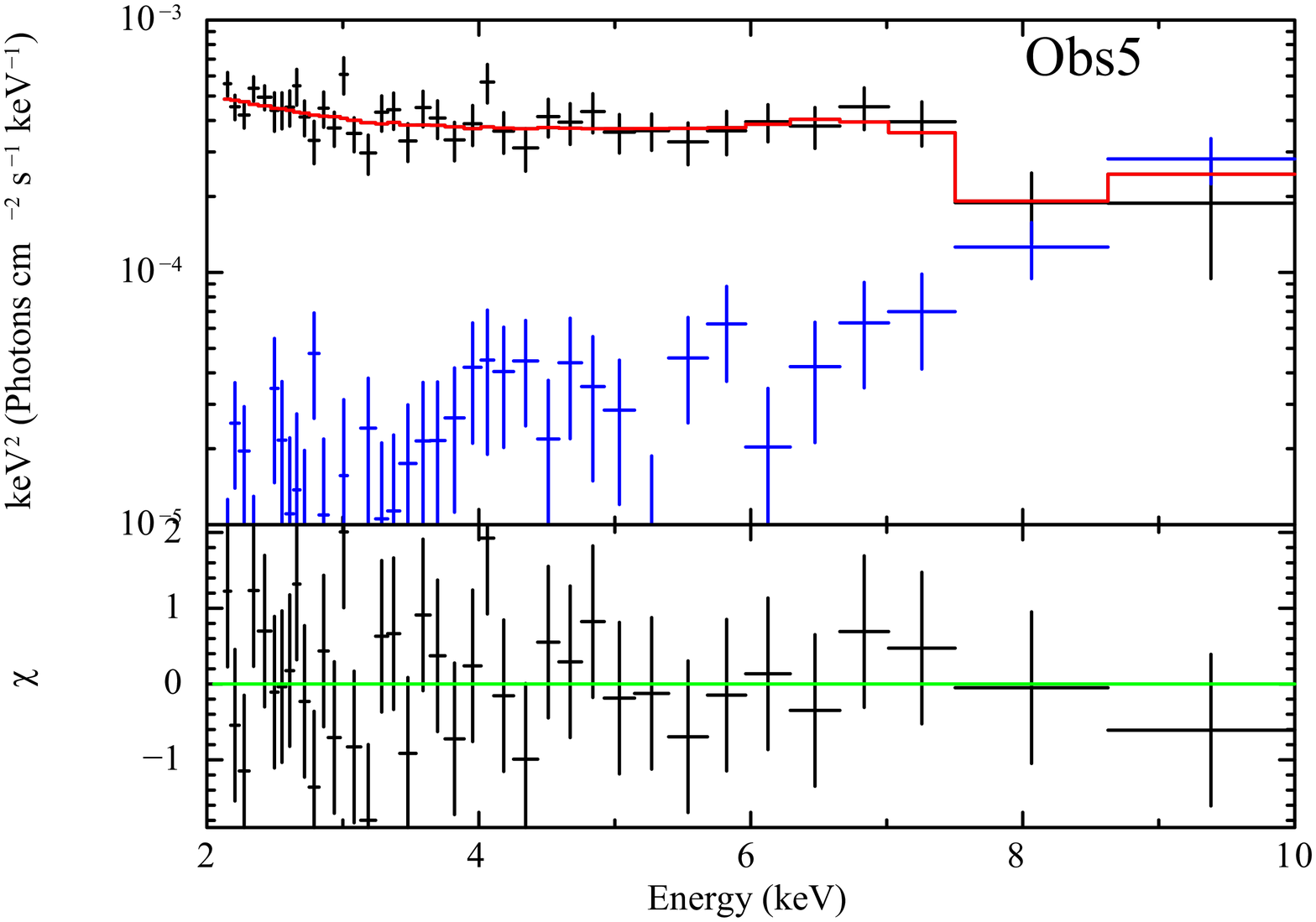}%6

\includegraphics[width=0.31\hsize]{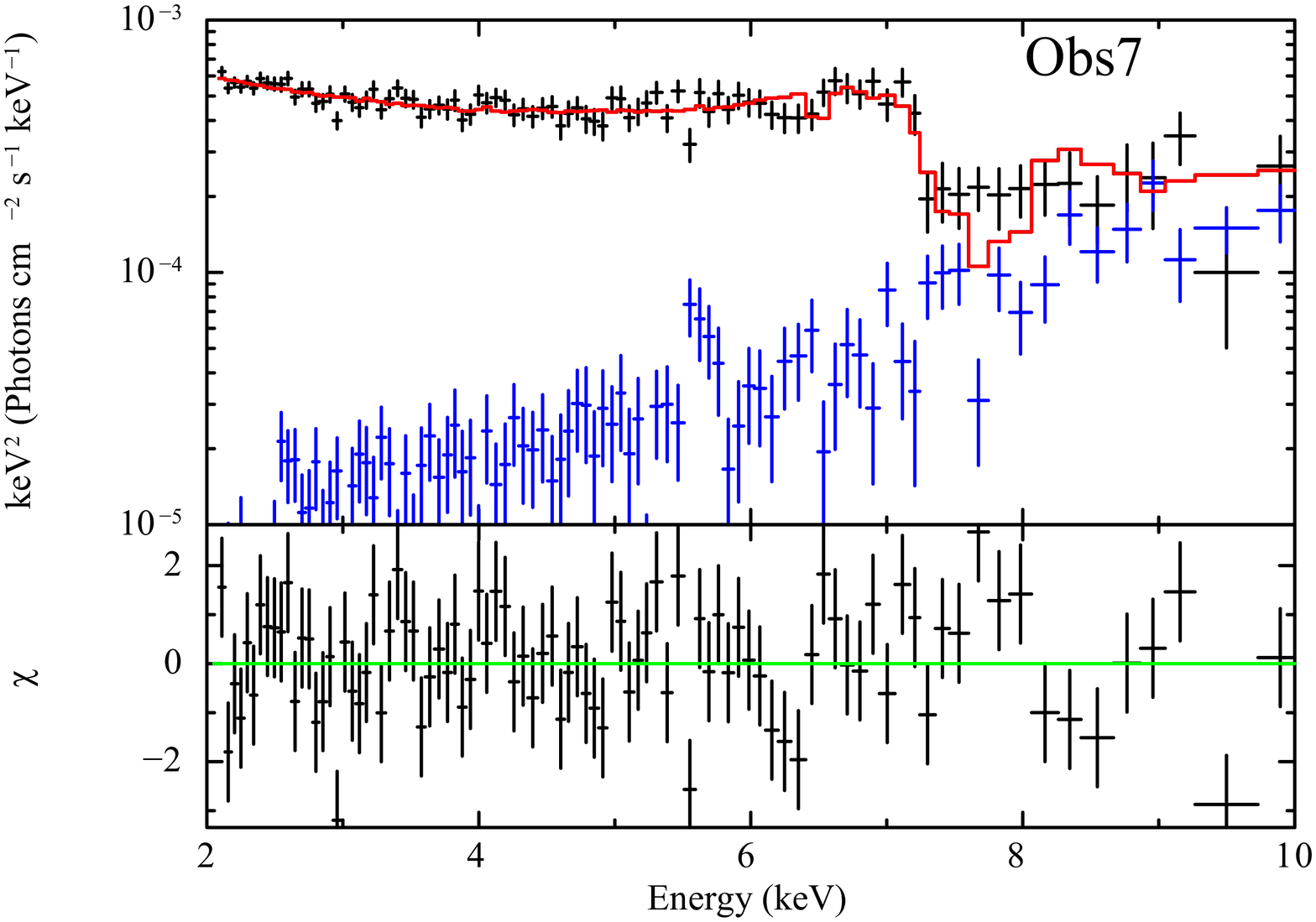}
\includegraphics[width=0.31\hsize]{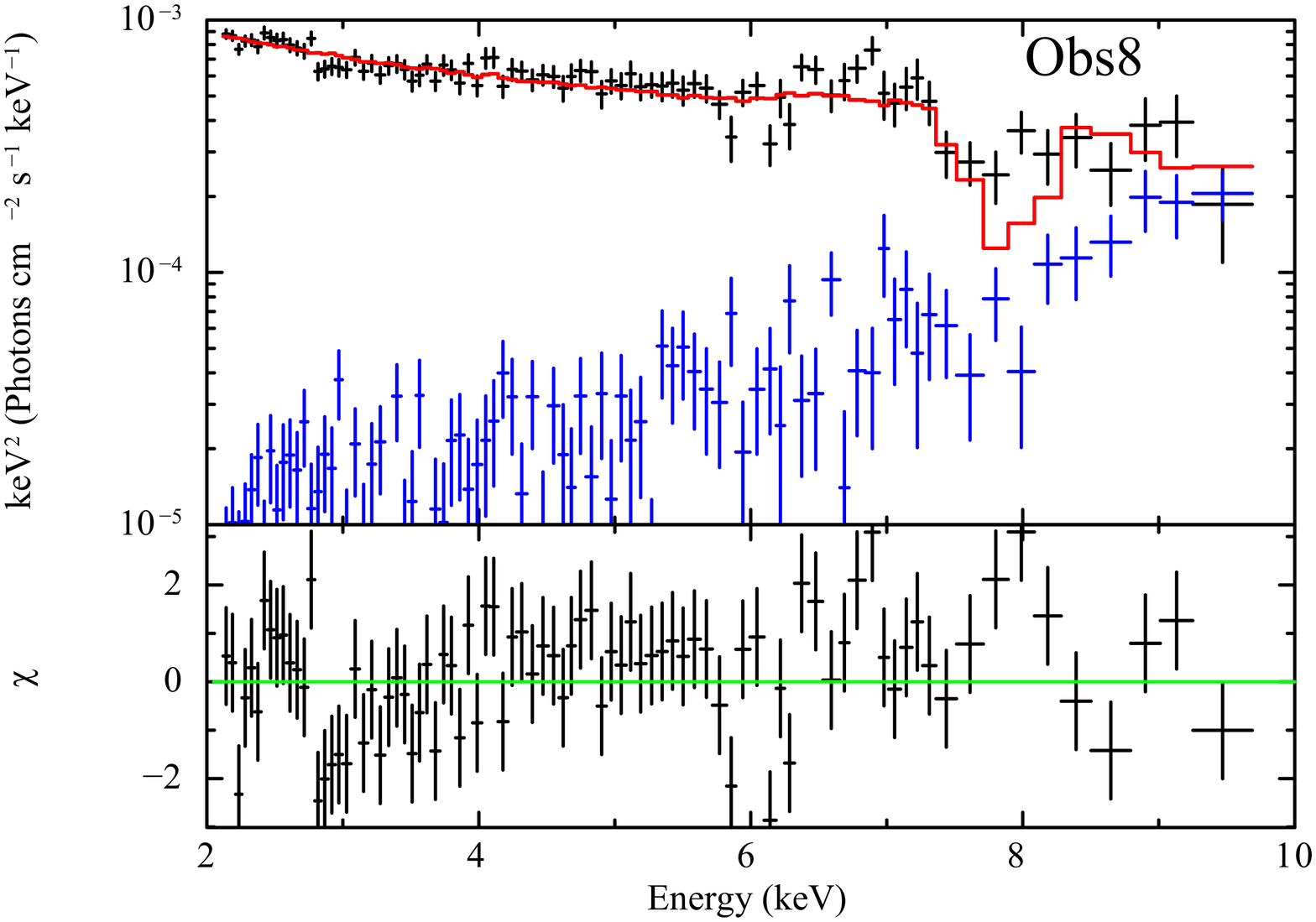}%8
\includegraphics[width=0.31\hsize]{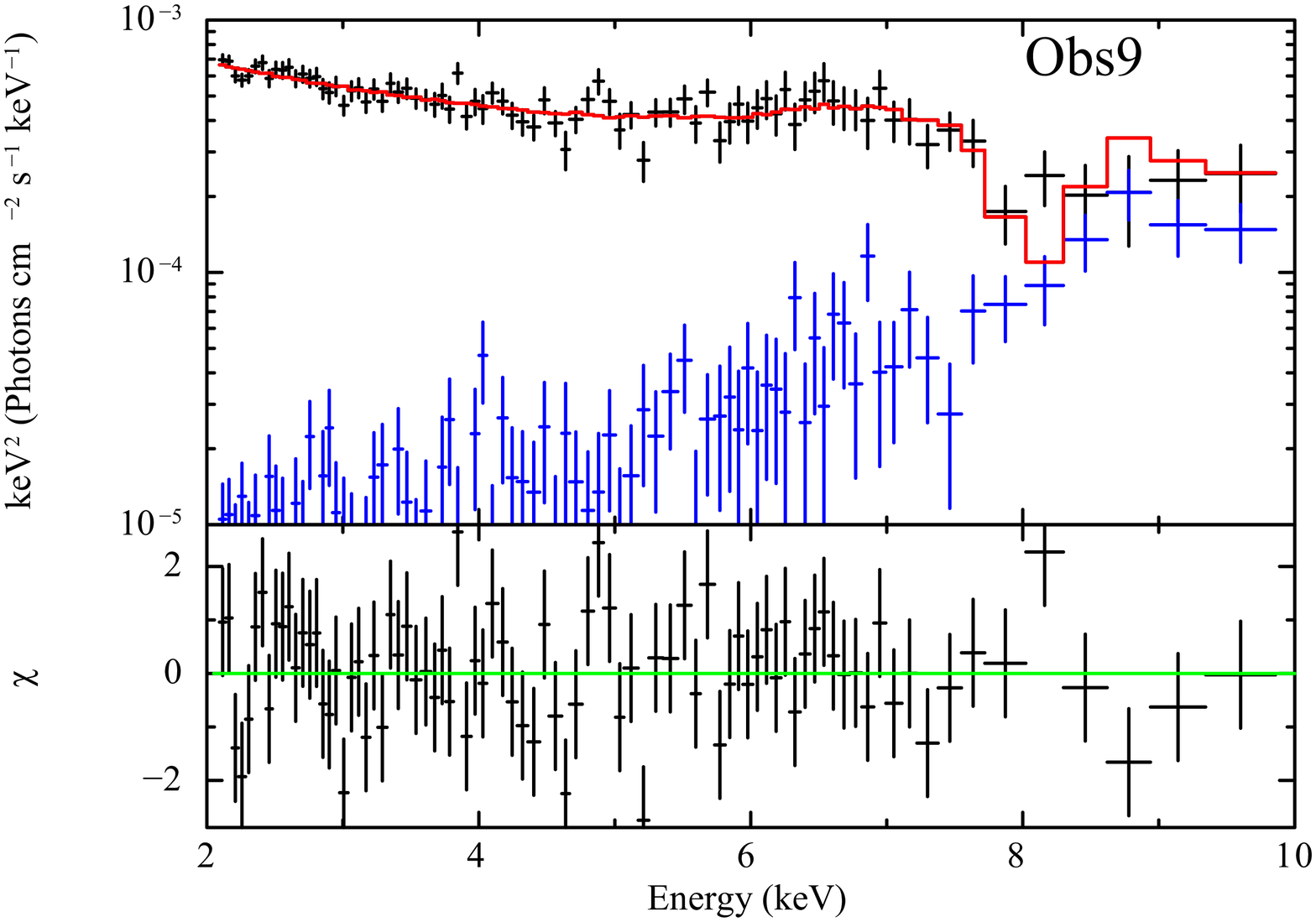}%9

\includegraphics[width=0.31\hsize]{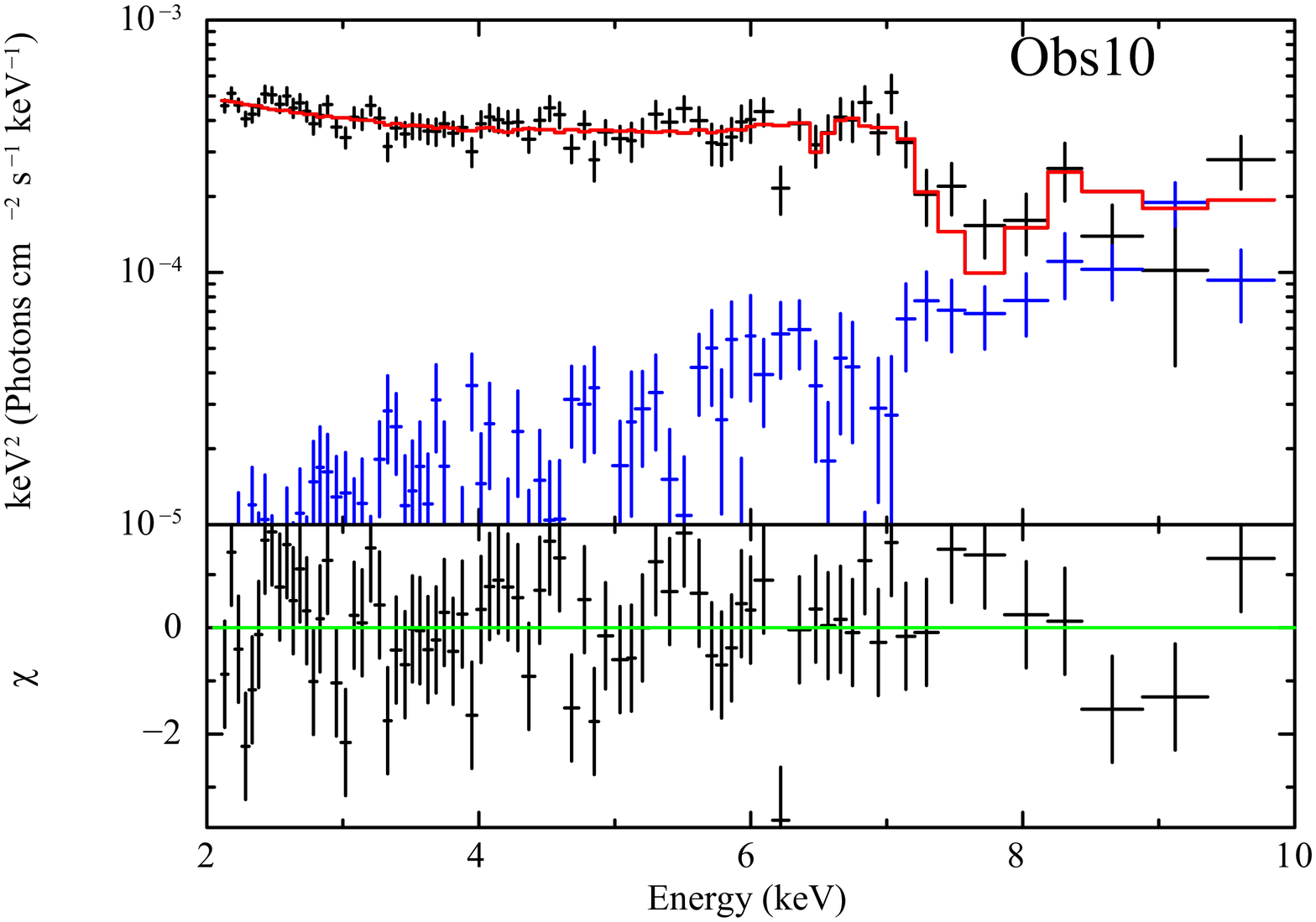}
\includegraphics[width=0.31\hsize]{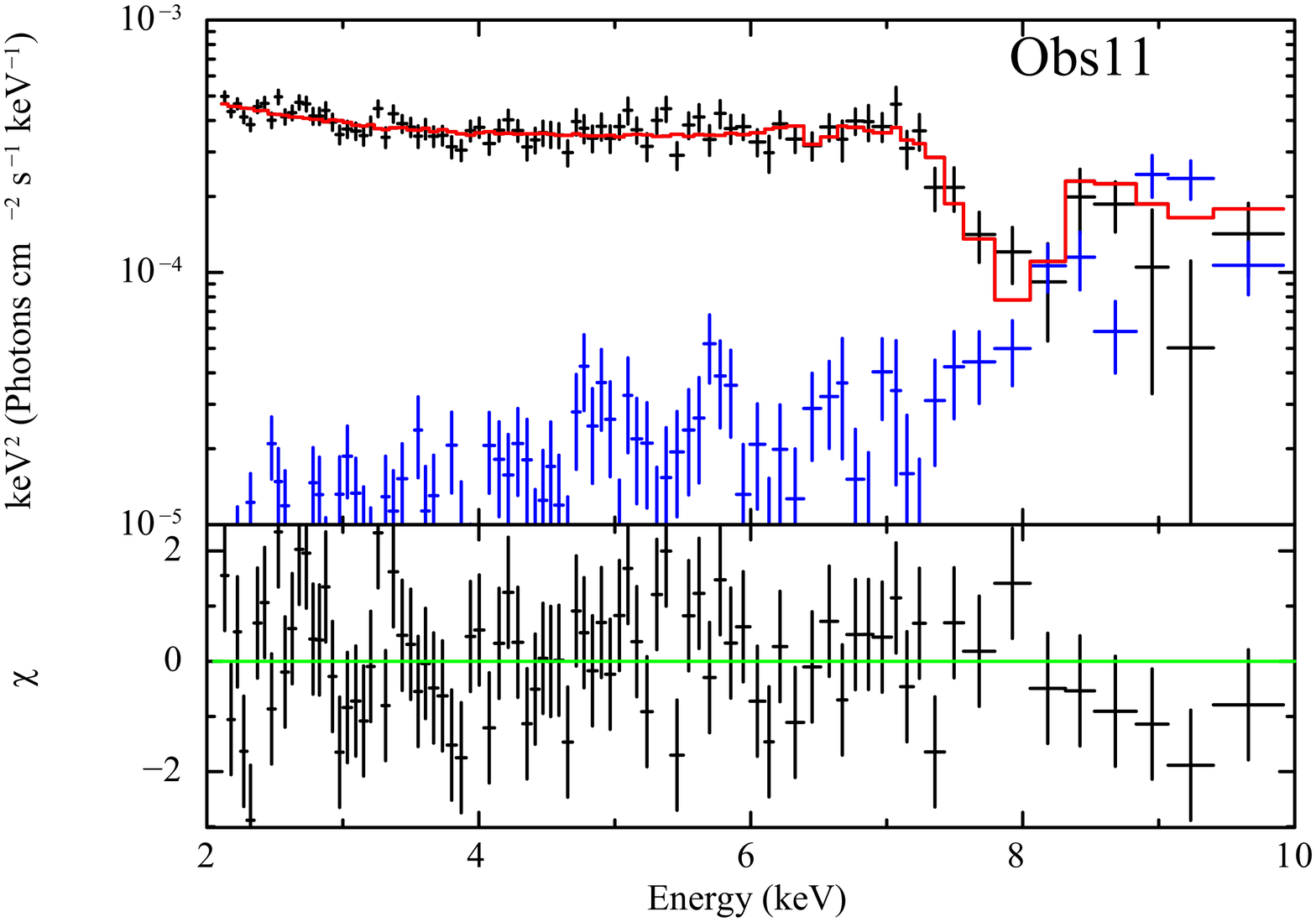}
\includegraphics[width=0.31\hsize]{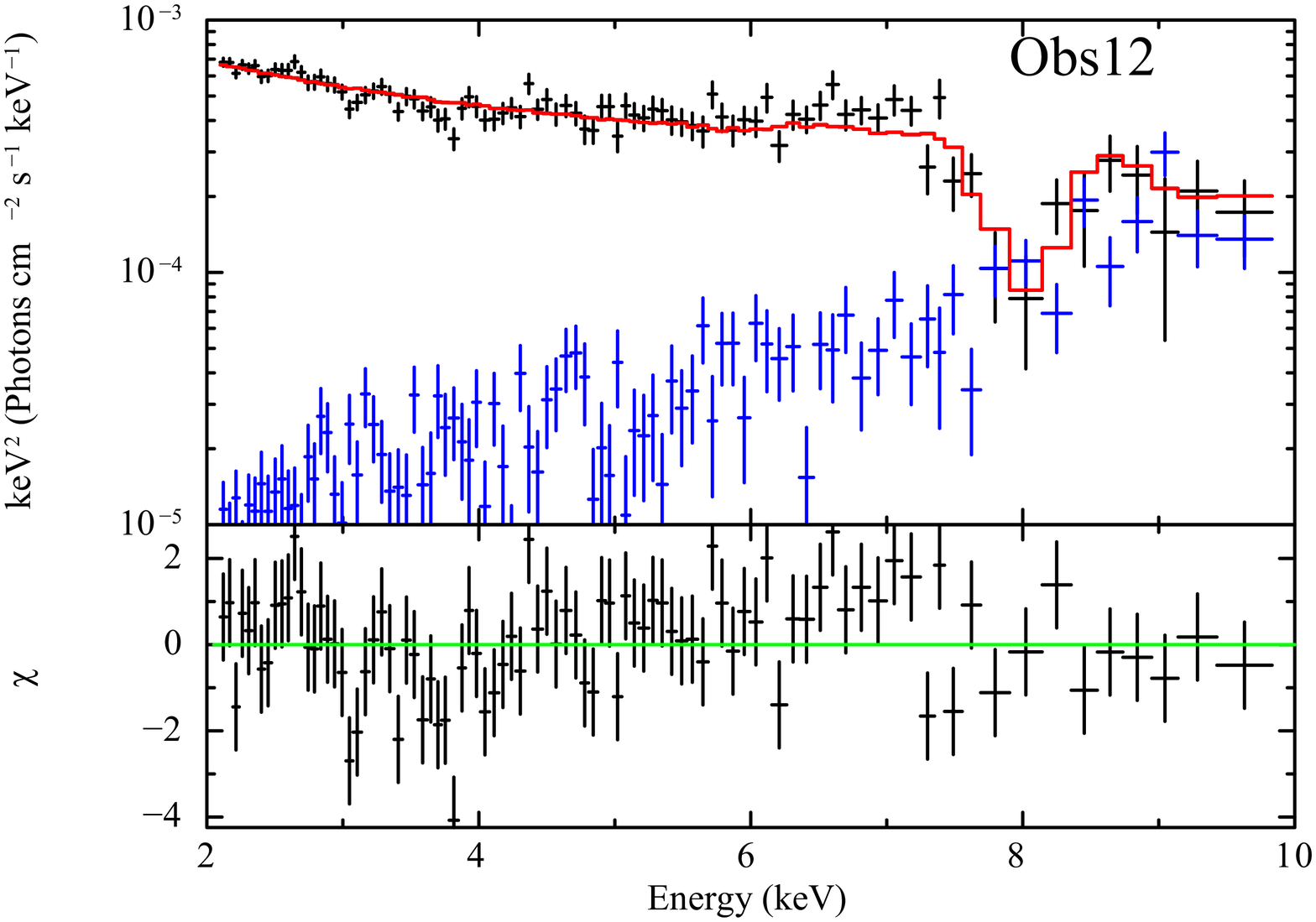}%12

\includegraphics[width=0.31\hsize]{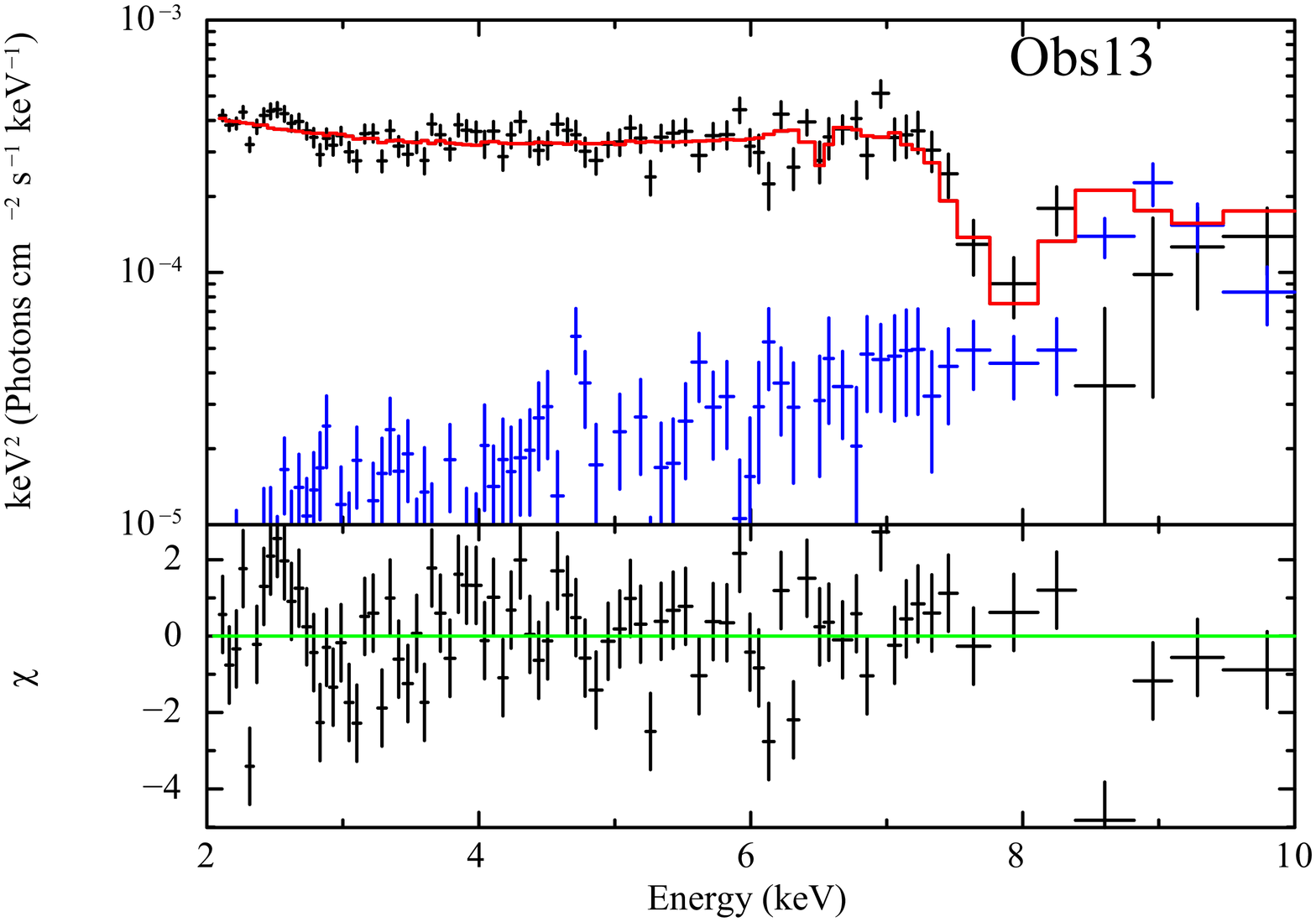}
\includegraphics[width=0.31\hsize]{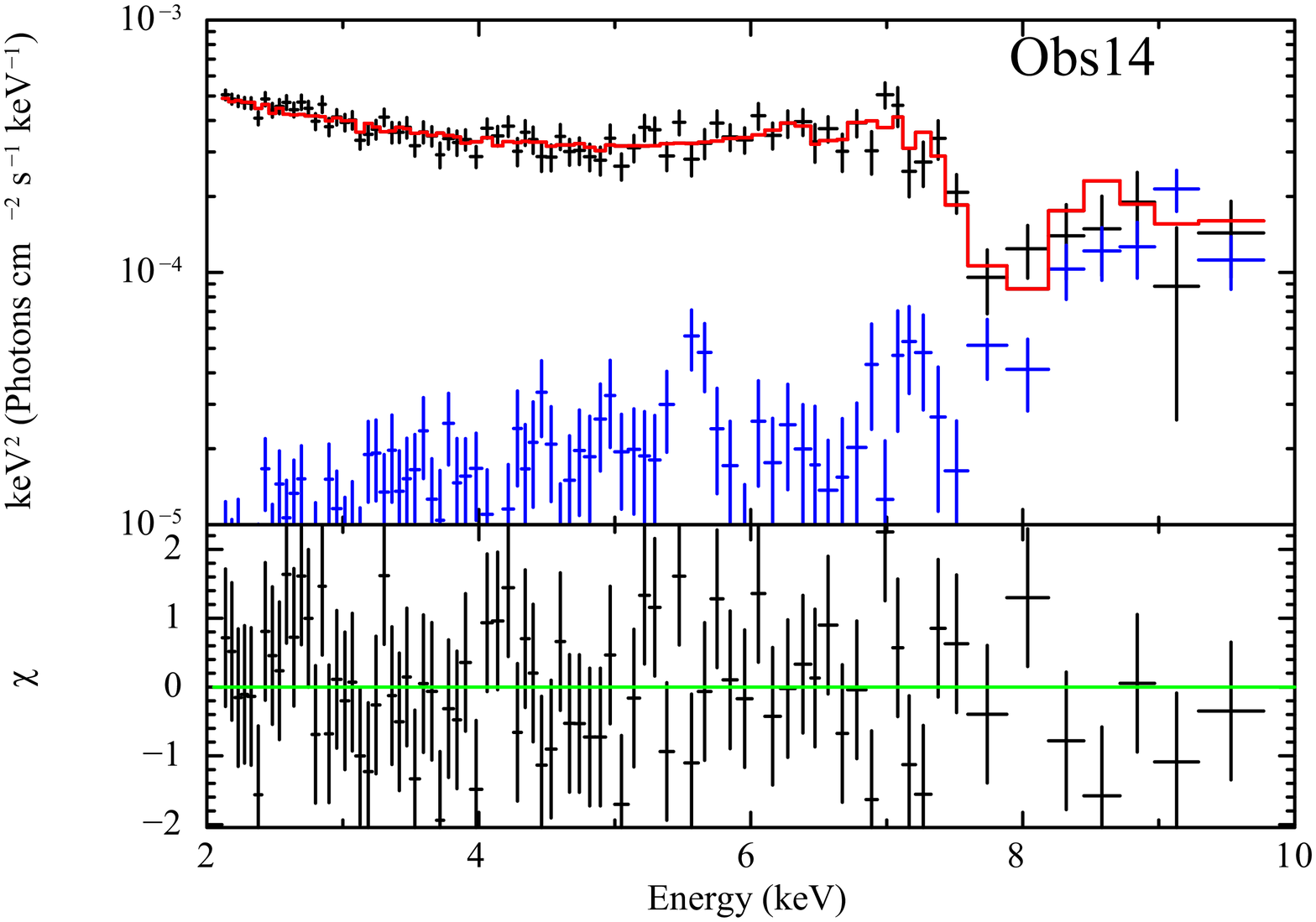}
\includegraphics[width=0.31\hsize]{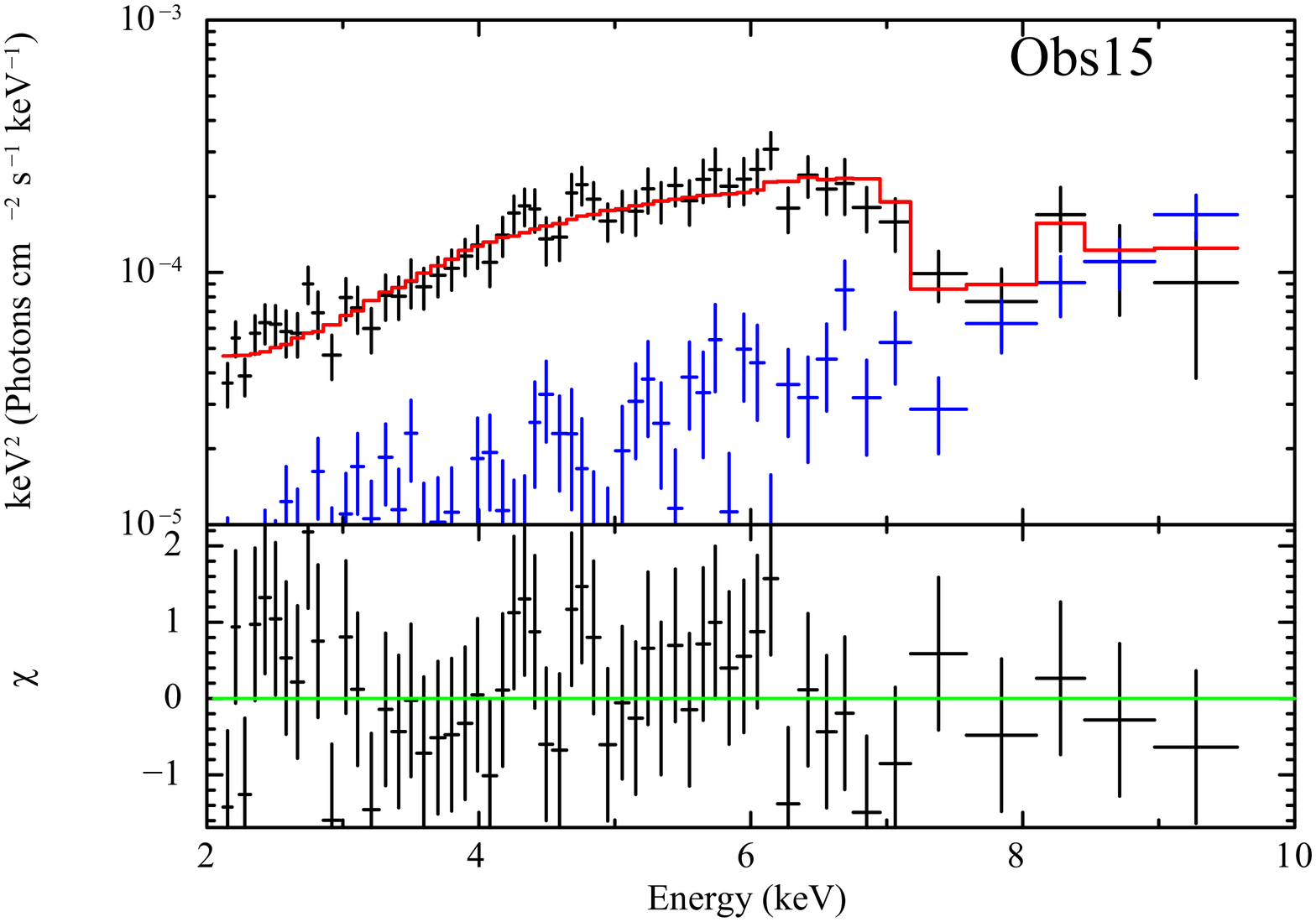}

\includegraphics[width=0.31\hsize]{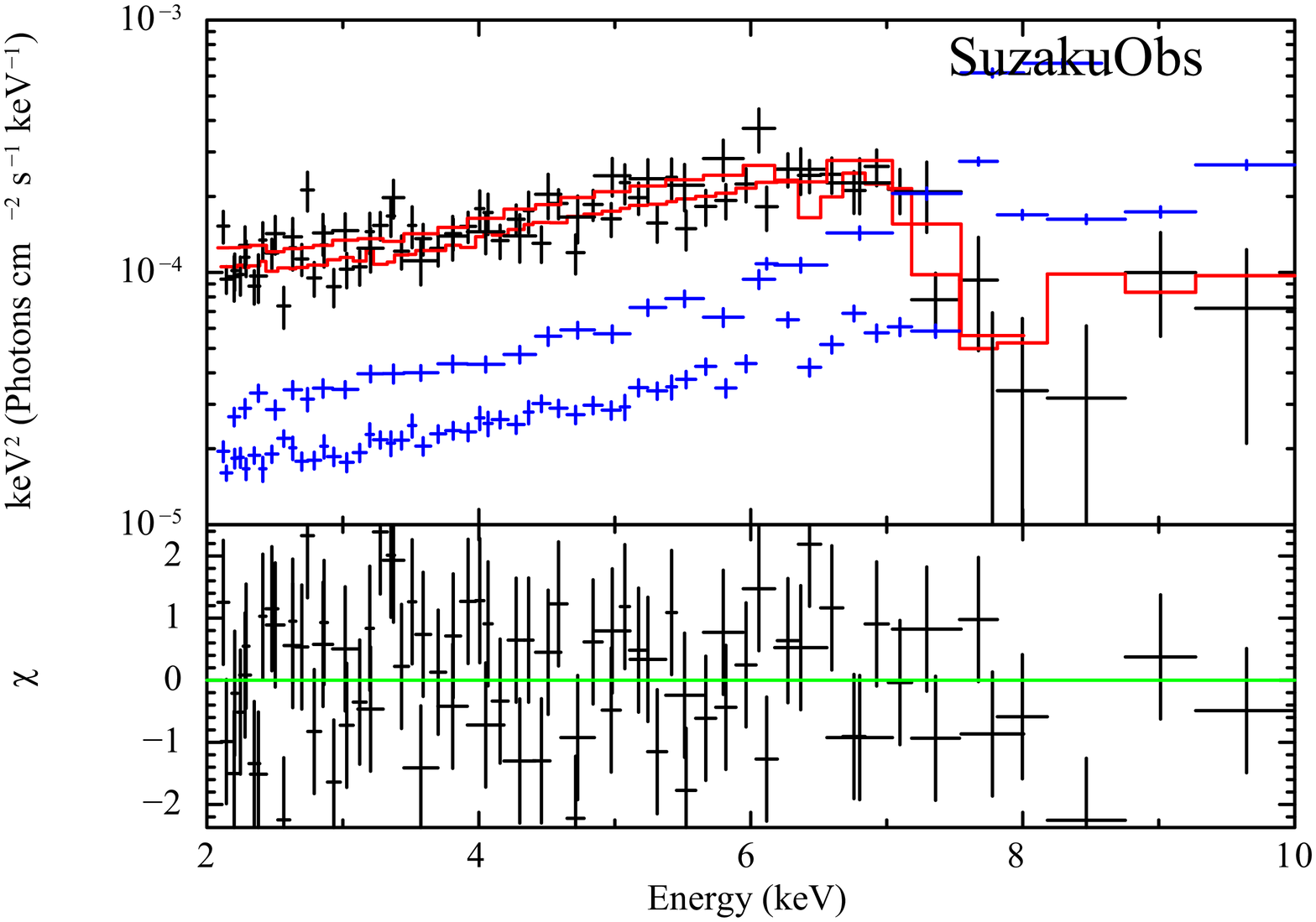}
\caption{Observed spectra and \monaco models with $v=0.2c$ and $\dot{M}_{\rm wind}/\dot{M}_{\rm Edd}=0.2$ for all the observations.
The spectra and models are plotted in black and red, respectively. The background spectra are also overplotted in blue. The lower panels show the residuals in units of $\chi$.
}
\label{fig:1H_fitspec_all}
\end{center}
\end{figure*}

All the spectra (black) and models (red) are plotted in
Fig.~\ref{fig:1H_fitspec_all}. We also show the background (blue) for each
spectrum, so that the signal-to-noise at high energies can be directly assessed. 
The parameters for each fit are listed in
Tab.~\ref{tab:1hparamall}. For Obs8 and Obs12, the 
absorption component is not included because the significance of
adding {\sc zxipcf} does not exceed the $99$\% confidence level of
an F-test.

As shown in Fig.~\ref{fig:1H_fitspec_all}, our \monaco models roughly
reproduce the sharp drop around 7.1--7.5~keV in all the observed
spectra. The best fit values of inclination angle ranges from
$\sim 58^\circ$ to $75^\circ$. These are all higher than the
inclinations inferred for \pds, and there is a broader spread in
derived inclination angle.

%\begin{landscape}
\begin{table*}
\caption{\monaco fit with $v=0.2c$ and $\dot{M}_{\rm wind}/\dot{M}_{\rm Edd}=0.2$ to all the observations of \hh.}
\begin{center}
\begin{tabular}{cccccccc}
\hline
\hline
Name & \multicolumn{3}{c}{Continuum absorption} & \monaco wind & \multicolumn{2}{c}{Fit statistics}\\
	  & $N_{\rm H}$ ($10^{22}$~cm$^{-2}$) & $\log\xi$ & $f_{\rm cov}$ & $\theta_{\rm incl}$ & $\chi^2$/dof & Null probability \\
\hline
Obs1 & $28_{-9}^{+83}$ & $1.90_{-0.63}^{+0.91}$ & $0.85_{-0.03}^{+0.12}$ & $74.6_{-5.4}^{+2.8}$ & $34.5$/$40$ & $0.73$ \\
Obs2 & $66_{-43}^{+72}$ & $1.90_{-0.87}^{+0.28}$ & $0.45_{-0.09}^{+0.45}$ & $64.1_{-1.9}^{+2.0}$ & $85.2$/$79$ & $0.30$ \\
Obs3 & $15_{-6}^{+37}$ & $<1.63$ & $0.68_{-0.05}^{+0.05}$ & $64.1_{-6.3}^{+2.8}$ & $52.7$/$55$ & $0.56$ \\
Obs4 & $8.2_{-6.3}^{+18.2}$ & $<2.28$ & $0.80_{-0.06}^{+0.19}$ & $57.9$ \footnotemark[1] & $15.1$/$21$ & $0.82$ \\
Obs5 & $23_{-16}^{+72}$ & $<2.14$ & $0.28_{-0.09}^{+0.44}$ & $68.0_{-1.6}^{+1.5}$ & $102.6$/$69$ & $0.005$ \\
Obs6 & $8.4_{-4.5}^{+90.6}$ & $<2.02$ & $0.42_{-0.14}^{+0.29}$ & $63.4_{-12.7}^{+5.6}$ & $28.2$/$34$ & $0.75$ \\
Obs7 & $151_{-95}^{+11}$ & $2.71_{-0.56}^{+0.10}$ & $0.71_{-0.25}^{+0.06}$ & $68.3_{-1.7}^{+1.7}$ & $115.1$/$87$ & $0.02$ \\
Obs8 & --- & --- & --- & $64.0_{-2.3}^{+2.5}$ & $129.12$/$84$ & $0.001$ \\
Obs9 & $65_{-31}^{+73}$ & $<2.26$ & $0.42_{-0.21}^{+0.41}$ & $58.3_{-1.9}^{+2.9}$ & $88.6$/$79$ & $0.22$ \\
Obs10 & $124_{-109}^{+33}$ & $2.71_{-1.90}^{+0.19}$ & $0.58_{-0.14}^{+0.16}$ & $68.2_{-2.6}^{+2.2}$ & $87.4$/$72$ & $0.10$ \\
Obs11 & $121_{-73}^{+32}$ & $2.76_{-0.46}^{+0.21}$ & $0.60_{-0.24}^{+0.20}$ & $63.1_{-2.6}^{+3.3}$ & $100.3$/$81$ & $0.07$ \\
Obs12 & --- & --- & --- & $60.6_{-1.8}^{+2.5}$ & $132.0$/$89$ & $0.002$ \\
Obs13 & $117_{-63}^{+28}$ & $2.73_{-0.31}^{+0.15}$ & $0.63_{-0.20}^{+0.13}$ & $63.2_{-2.6}^{+2.4}$ & $154.9$/$78$ & ${5\times10^{-7}}$ \\
Obs14 & $181_{-8}^{+203}$ & $2.82_{-0.09}^{+0.10}$ & $0.82_{-0.08}^{+0.07}$ & $63.1_{-2.6}^{+2.3}$ & $75.3$/$77$ & $0.53$ \\
Obs15 & $23_{-6}^{+47}$ & $1.61_{-0.78}^{+1.02}$ & $0.93_{-0.02}^{+0.04}$ & $70.9_{-3.8}^{+3.7}$ & $43.1$/$51$ & $0.78$ \\
SuzakuObs & $109_{-69}^{+27}$ & $2.72_{-1.47}^{+0.07}$ & $0.89_{-0.08}^{+0.08}$ & $67.0_{-5.4}^{+5.9}$ & $97.9$/$86$ & $0.18$ \\
\hline
\end{tabular}
\end{center}
\begin{flushleft}
\footnotesize
$^1$ No constraints on this parameter were obtained.
\end{flushleft}
\label{tab:1hparamall}
\end{table*}%
%\end{landscape}

%\clearpage

\subsection{Application to \nustar data}

A key breakthrough in AGN spectral studies has come from \nustar data
which extend the energy range of the observed spectra beyond 10~keV
and have much better signal-to-noise above 7~keV than the \suzaku or \xmm data
(see Fig. \ref{fig:1H_fitspec_all}). 
\hh has also been observed by \nustar, and the resulting spectra can be
fit by the extreme relativistic reflection models \citep{Kara2015}.

\begin{figure}
\begin{center}
\includegraphics[width=\hsize]{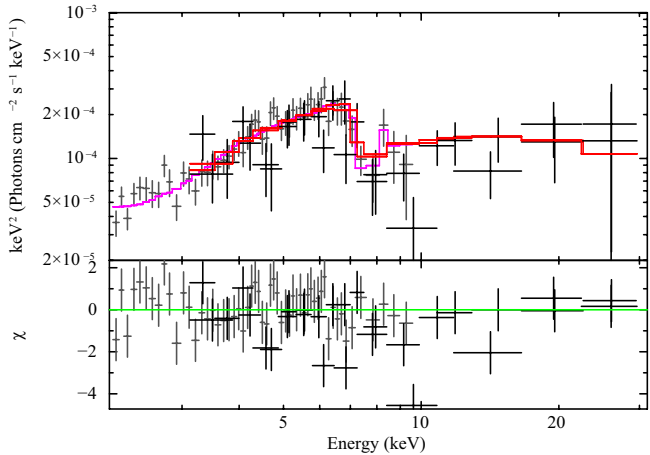}
\caption{\nustar data compared with the best-fit wind model of Obs15.
The spectra of \xmm, \nustar are plotted in grey and black, while the model lines for \xmm and \nustar are plotted in magenta and red, respectively. The lower panels show the residuals in units of $\chi$.
}
\label{1h_nustar}
\end{center}
\end{figure}

We extract the \nustar data, and follow \cite{Kara2015}
in selecting the second \nustar dataset, which is a good match to \xmm
Obs15. Fig.~\ref{1h_nustar} shows the model fit to Obs15 {\em
extrapolated} to 30~keV.  We note that
around half the drop at $\sim 7$ keV is from the wind model alone,
while the other half is from the complex lower ionization absorption.
There are no additional free parameters, but
the model gives a good fit to the higher energy data. 

\section{Discussion}\label{sec:discussion}

\subsection{Effects of cool clumps}
We model the wind using the \monaco Monte-Carlo code, which tracks
emission, absorption and scattering in a continuous wind geometry
\citep{Hagino2015}. In this continuous wind model, all the atoms
lighter than iron are almost fully ionized, so that it cannot produce
the strong spectral variability seen below the iron line region.
Instead, in the archetypal wind
source \pds, such variability is assumed to be from lower ionization
species. In \pds this
additional spectral variability can be approximately modeled using
partially ionized material which can partially cover the source. We
show that this same combination of partial covering by lower
ionization material together with the highly ionized wind gives an
acceptable fit to the \hh spectra, including the higher energy data
from \nustar.

The lower ionization material in the wind is likely to be
clumped due to the ionization instability for X-ray illuminated material
in pressure balance \citep{Krolik1981}. These clumps are cooler and
less ionized than the hot phase of the wind, but with lower filling factor.
The dramatic dips which are characteristic of the light curves of complex
NLS1 like \hh can then be interpreted as occultations by these cool clumps.
We note that time dependent, clumpy absorption is typical of both
UV line driven disk winds \citep{Proga2004}, and continuum driven
winds \citep{Takeuchi2014}.

\begin{figure}
\begin{center}
\includegraphics[width=\hsize]{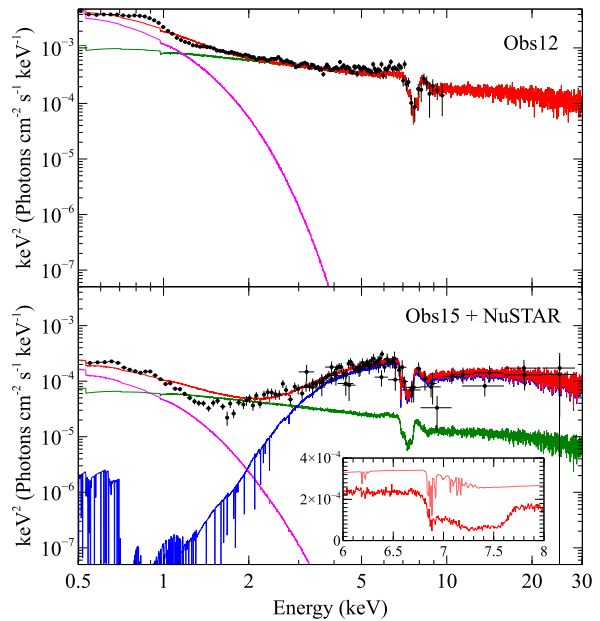}
\caption{The observed data and model of the highest flux state (Top: Obs12)
and the lowest flux state (Bottom: Obs15 and the \nustar data) plotted over
a wide energy range. The absorbed spectrum is shown as the blue line,
the unabsorbed in green, the total in red, and the observed data in black.
A soft X-ray excess component with typical parameters is also plotted in magenta to show
the spectrum at lower energies below 2~keV. There are no spectral features
predicted below 2~keV due to the dilution by the unabsorbed
(or reflected/scattered) flux (green). There are absorption lines from
Fe~\textsc{xviii}--\textsc{xx} produced by cool clumps (pink line in a small
window inside bottom panel) but the K$\alpha$ lines at $\sim6.45$~keV are
not detectable even with 250~ks of {\it Hitomi}, and the higher energy lines
merge with the hot phase of the wind (red line in a small window).}
\label{cool}
\end{center}
\end{figure}

The partially ionized clumps which partially cover the source imprint
some spectral features as well as curving the continuum. Fig.~\ref{cool}
shows the data and model in the entire energy band for the highest
(top panel) and lowest (bottom panel) flux states. The data is plotted
in black and the total model spectrum is shown by the red line. The model
spectrum is separated into absorbed (blue), unabsorbed
(green) and a soft X-ray excess component (magenta, see below).
The low ionization absorption imprints strong atomic features from
oxygen and iron-L below 2~keV in the lowest intensity state, but these
are diluted by the X-rays which are not absorbed, so that
the total spectrum is almost featureless in the 0.5--2~keV bandpass.
Hence it predicts no observable features in the RGS.

The unabsorbed component does not dilute the iron-K features
because the absorbed component is dominant at higher energies.
The effect of absorption at iron K from the cool clumps alone is shown by the pink
line in the inset window of Fig.~\ref{cool}, while the red line in the inset
shows the total (cool plus hot phase) absorption. The pink line shows
that the cool clumps do imprint iron K$\alpha$ absorption lines around
6.45~keV rest frame from Fe~\textsc{xviii}--\textsc{xx} as well as
the much stronger K$\beta$ lines around 7~keV rest frame. The K$\beta$
lines blend with the hot phase absorption, while the K$\alpha$ lines have
an equivalent width of only a few eV, which is not detectable even with a
300~ks observation with {\it Hitomi} for a source as weak as \hh. Hence
the model is consistent with current and near future limits on spectral
features in the data.

A soft excess component (magenta) is added in the model spectrum in
Fig.~\ref{cool} in order to roughly describe the spectrum 
at low energies. We use a thermal Comptonization model {\sc comptt},
allowing only the normalization to be free while fixing the shape to
parameters typically seen in other AGN, namely a seed photon
temperature of 0.05~keV, coronal temperature of $kT = 0.2$~keV and
optical depth at $\tau=15$ \citep{Done2012}. The plots show that the
observed spectrum is roughly reproduced by this model, but there is an
excess around 0.8~keV and a deficit between $\sim1$--2~keV. These
differences can be interpreted as broad emission and absorption (P Cygni profile)
from oxygen K and iron L-shell transitions, which are naturally expected in 
the ionized fast wind, but are not included in our current Monte-Carlo code. 
Hence these features in the data are not well described by the model used here, which is tailored instead to
iron K. Nonetheless, we do expect that similar features would be produced
by including emission and absorption from cool clumps in our wind model.

Our model is currently incomplete as these clumps should also
reflect/emit as well as absorb, adding to the reflection/emission from
the highly ionized phase of the wind which is included in our \monaco
simulation. This reflection/emission from the composite wind could
dominate during the dips, where the cool clumps probably completely
cover the source itself. We also envisage the cool clumps to be
entrained in the wind, so the atomic features should also be
blueshifted/broadened by the wind velocity structure. This means that
the cool phase of the wind also 
contribute to the total wind kinetic luminosity which has so far been
neglected.

The clumpy structure can also play an important role for a
characteristic time lags observed in \hh.  The observations show a
complex pattern of lags between soft and harder energies, with the
soft band leading for slow variability, and lagging for fast
variability \citep{Fabian2009,Zoghbi2010}. In the relativistic reflection picture,
this can be explained by a partially ionized disk, where reflection is
weaker at 2--4~keV than at lower or higher energies. Thus, reflection
dominates in the soft band, while the intrinsic continuum dominates in
the intermediate band at a few keV. Reflection follows the continuum,
but with a lag from the light travel time from the source to the
reflector, giving the soft (conventionally referred to as a `negative')
lag at high frequencies. The positive lag at lower frequencies
is probably from propagation of fluctuations through the accretion
flow. On the other hand, a full spectral-timing model including both the positive and
negative lags for PG~1244+026, another AGN with very similar mass and
mass accretion rate to \hh, gives the observed lag of 200~s seen in
PG~1244+026 from reprocessing by material extending from 6--12$R_{\rm g}$
\citep[assuming a mass of $10^7$\msun:][]{Gardner2014}.
In this interpretation, this lag is due to the illuminating hard X-ray flux which
is not reflected but thermalized and re-emitted as quasi-blackbody radiation.
Adding multiple occultations by clumps onto the same full spectral-timing
model reduces the predicted lags to 50~s, similar to the 30~s observed in
\hh but without requiring a small reflector distance
\citep{Gardner2015}.  Thus cool clumps may be able to explain both the
time averaged spectrum and the extremely short lags without requiring
lightbending from a small corona close to the event horizon of an
extreme spin black hole, though it remains to be seen whether they can 
fit the lag-energy spectra as well as the lag frequency spectra
\citep{Gardner2015}. However, we note that the lamppost model itself also has 
difficulty fitting the details of the lag-energy spectra
\citep{Wilkins2016}. Their alternative geometry of an extended corona may give a 
better fit to the lag-energy spectra, but it seems unlikely to be able to simultaneously
explain the deep dips in the light curves which can be explained in the lamppost model 
by the very small source approaching the horizon.

\subsection{Wind geometry}

Our specific wind model is probably not unique in terms of wind
geometry and velocity structure. In particular, we used the same wind
solid angle of $\Omega/4\pi=0.15$ in \hh as for \pds.
This is likely to be appropriate for \pds if there is indeed some component
of UV line driving to the wind. It is because numerical studies show that
this driving mechanism results in time variable but fairly narrow wind streamlines.
However, continuum radiation driving is much more likely in \hh, in which case the
numerical simulations predict a wider opening angle wind. Future work
should simulate winds with a larger opening angle, which would require
correspondingly larger mass loss rates in order to keep the same
absorption column density.

The wider opening angle wind should also result in stronger line emission,
so that this may give a better fit to some of the remaining residuals.
However, we caution that all these spectra are co-added,
integrating over dramatic short timescale variability so steady state
models may not be appropriate. 

\section{Conclusions}
We can successfully reproduce the range of spectra seen from \hh with
absorption in a wind from the inner disk.
The strong spectral drop is produced by our line of sight cutting
across the acceleration region where the wind is launched.
In this region, we see the absorption line over a wide range of velocities,
making a very broad feature. Since the column density in the wind is large,
the broad absorption line merges into the absorption edge, which further
depresses the spectrum at higher energies.
%A strong wind from the inner disk is expected in \hh since its optical/UV spectrum implies that the accretion flow is strongly super Eddington \citep{Done2015}.

Our wind model probably inappropriate to this source as discussed in section \ref{sec:discussion}.
The time variability and the low ionization absorption in the observed spectra
seems to be consistent with an existence of cool clumps in the wind.
This clumps are naturally expected due to the instability of the wind.
Also, a larger opening angle of the wind would probably give better fits to the observed data.
This implies the continuum radiation driving mechanism, which is expected in the super Eddington accretion.

Nonetheless, the wind model presented here can fit the overall 2--30~keV
spectra from \hh. Unlike the lamppost models, it does not require
that the black hole has extreme spin, nor does it require that we have
a clean view of a flat disk, nor does it require that iron is
7--20 times overabundant. We suggest that the extreme iron features
in \hh arise more from absorption/scattering/emission from an inner
disk wind than from extreme gravity alone.

\section*{ACKNOWLEDGMENTS}

K.H. was supported by the Japan Society for the Promotion of Science
(JSPS) Research Fellowship for Young Scientists. CD acknowledges STFC
funding under grant ST/L00075X/1. This work was supported by
JSPS KAKENHI grant number 24740190 and 15H06897.

\bibliographystyle{mnras}
\bibliography{ref}

\label{lastpage}
\end{document}